\documentclass[12pt]{article}

\usepackage[margin=1in]{geometry}
\usepackage{amssymb,amsmath,latexsym}                        
\usepackage{graphicx}
\usepackage{cite}
\usepackage{comment}
\usepackage{xcolor}
\usepackage{float}

\usepackage{amsthm}
\usepackage{multicol}
\usepackage[colorlinks=true, linktoc=page, pdftex, linkcolor=blue, citecolor=blue, urlcolor=blue]{hyperref}
\usepackage{adjustbox}
\usepackage{bbm}
\usepackage{mathtools}

\usepackage{adjustbox}

\date{\today} 

\begin{document}

\title{\vspace{-2cm}{\normalsize\hspace{120mm}\vspace{10mm}CERN-TH-2017-184}\\
Observables for possible QGP\\ signatures in central $pp$ collisions}
\author{
Benjamin Nachman$^{a}$ and Michelangelo L. Mangano$^{b}$ \\
{\normalsize\it $^{a}$Physics Division, Lawrence Berkeley National Laboratory, Berkeley, CA 94704, USA}\\
{\normalsize\it $^{b}$Theoretical Physics Department, CERN,  1211 Geneva 23, Switzerland}}

\maketitle

\begin{abstract}
Proton-proton ($pp$) data show collective effects, such as long-range azimuthal correlations and strangeness enhancement, which are similar to phenomenology observed in heavy ion collisions.  Using simulations with and without explicit existing models of collective effects, we explore new ways to probe $pp$ collisions at high multiplicity, in order to suggest measurements that could help identify the similarities and differences between large- and small-scale collective effects.  In particular, we focus on the properties of jets produced in \textit{ultra-central} $pp$ collisions in association with a $Z$ boson. We consider observables such as jet energy loss and jet shapes, which could point to the possible existence of an underlying quark-gluon plasma, or other new dynamical effects related to the presence of large hadronic densities. 
\end{abstract}

\section{Introduction}
\label{sec:intro}

There has been a recent surge of interest in collective effects in small systems with high final state multiplicity due to measurements of strangeness enhancement from ALICE~\cite{ALICE:2017jyt} and large-angle particle correlations (the `ridge') by ATLAS~\cite{Aad:2015gqa} and CMS~\cite{Khachatryan:2010gv,Khachatryan:2015lva}.  These effects are not reproduced by the standard Monte Carlo (MC) event generators for $pp$ collisions, based on standard Quantum Chromodynamics (QCD) evolution and well-tested models of hadronization~\cite{Sjostrand:2006za,Corcella:2000bw,Bahr:2008pv,Gleisberg:2008ta,Sjostrand:2007gs,Sjostrand:2014zea,Bellm:2015jjp}. The features of these phenomena resemble those exhibited by the Quark Gluon Plasma (QGP) formed in heavy ion  (HI) collisions. If parameterized in terms of $dN/d\eta$, the evolution of the observed effects with $d\text{N}/d\eta$ in $pp$ smoothly matches to the size of the effects observed in HI collisions, where they are interpreted in terms of QGP dynamics (see e.g. Fig. 2 in Ref.~\cite{ALICE:2017jyt}). It is therefore tempting to speculate that a sort of ``mini-QGP'' might be formed in (or might be responsible for) the highest $dN/d\eta$ events in $pp$.  Alternative interpretations have nevertheless been put forward, relying on a more complex description of the fragmentation phase of the event generation~\cite{Bierlich:2014xba,Pierog:2013ria,Fischer:2016zzs}. These descriptions of the collective phenomena make no reference to a QGP, and derive their results from a more extended network of interactions among the partons emerging from the usual ($T=0$) evolution of the partonic final state. More generally, the experimental facts raise the question of whether the description of large-multiplicity final states in $pp$ collisions boils down to finding the right knobs to tune in some fragmentation model, or whether it requires the understanding of a new dynamical phase of high-energy hadronic interactions.  

In this paper we propose a set of observables that, while being sensitive to the reported collective effects, would likely lead to different results depending on whether the QGP is active or not. In particular, we consider jet observables, which in the presence of a QGP are expected to undergo quenching effects that may not exist in non-QGP models of collective effects in $pp$. We analyze $Z$+jet events, and study the properties of the jets and of the surrounding environment, as a function of the track multiplicity. We focus on both the strangeness enhancement and on the potential quenching of the jet recoiling against the $Z$ boson. We show that the MC models predicting strange enhancement in high-multiplicity minimum bias events continue exhibiting large differences in the modeling of strange hadron production, with respect to the standard MCs. We also show, perhaps not surprisingly, that those models do not lead to an observable quenching of the jet energy, and an observable such as $p_\text{T,$J$}/p_\text{T,$Z$}$ shows no significant dependence on $dN/d\eta$, matching the prediction of MCs that do not model collective effects. 

We suggest that the experimental study of strangeness enhancement and quenching in $Z$ (or $\gamma$) plus jet events might help in better assessing the true nature of the collective phenomena recently observed in $pp$ collisions, proving or disproving their QGP-like origin, and providing valuable data to improve the MC models and their tuning.  Studies in $pPb$ by ATLAS~\cite{Aad:2012gla,Aad:2013fja,Aad:2014lta,ATLAS:2014cpa,Aad:2015ddl,Aad:2016zif,Aaboud:2016jnr,Aaboud:2016yar,Aaboud:2017xpw,Aaboud:2017acw,Aaboud:2017tke,Aaboud:2017blb}, CMS~\cite{Sirunyan:2017mzd,Sirunyan:2016fcs,Khachatryan:2016odn,Khachatryan:2016got,Khachatryan:2016yru,Khachatryan:2016ibd,Khachatryan:2015sva,Khachatryan:2015uja,Khachatryan:2015oea,Khachatryan:2015waa,Chatrchyan:2014hqa,Chatrchyan:2013nza,Chatrchyan:2013eya,Chatrchyan:2013nka,CMS:2012qk}, ALICE~\cite{ALICE:2012xs,ALICE:2012mj,Abelev:2012ola,Abelev:2013bla,Abelev:2013haa,Abelev:2013yxa,Abelev:2014pja,Abelev:2014dsa,Abelev:2014hha,Abelev:2014zpa,Abelev:2014mda,Abelev:2014mva,TheALICE:2014dwa,Abelev:2014oea,Adam:2014qja,Adam:2015pya,Adam:2015hoa,Adam:2015xea,Adam:2015iga,Adam:2015bka,Adam:2015jsa,Adam:2015jca,Adam:2015gda,Adam:2015qda,Adam:2015vsf,Adam:2015pbc,Adam:2016dau,Adam:2016bpr,Adam:2016mkz,Adam:2016ohd,Adam:2016jfp,ALICE:2016clc,Adam:2016ich,Adam:2016wyz,Adamova:2017elh,Acharya:2017goa,Adamova:2017opl,Adamova:2017uhu,Acharya:2017ino}, and LHCb~\cite{Aaij:2013zxa,Aaij:2014mza,Aaij:2015qcq,Aaij:2016eyl,Aaij:2017cqq,Aaij:2017gcy} have also started to probe an intermediate regime between $pp$ and $PbPb$, though with limited datasets compared with $pp$.  

This paper is organized as follows.  Section~\ref{sec:simulation} describes the simulation setup before the results are presented in Sec.~\ref{sec:results}.   Inclusive strangeness enhancement is demonstrated in Sec.~\ref{sec:strangeness} in addition to probing strangeness inside jets.  Momentum balancing is investigated in Sec.~\ref{sec:jetbalance} and additional observables related to jet substructure are studied in Sec.~\ref{sec:jetstructure}.  We present our conclusions in Sec.~\ref{sec:concl}.

\section{Simulation}
\label{sec:simulation}


\subsection{Collective Effects}
\label{sec:coleffects}

There have been several attempts within the context of the string hadronization model~\cite{Andersson:1983ia} to describe collective effects.  In string hadronization, color flux tubes (`strings') connecting partons produce a linearly confining potential.  Kinks in the string describe gluons and strings can break into $qq'$ pairs; various phenomenological and physics-inspired parameters determine the distribution of hadrons produced.

One very promising extension that provides an excellent description of strangeness enhancement observed by ALICE~\cite{ALICE:2017jyt} is the rope hadronization model.  Rope hadronization extends the string picture by allowing nearby strings to interact coherently to have an effectively higher string tension.  Due to their larger masses, strange hadrons are suppressed with respect to purely $u$/$d$ hadrons by $\sim m_s^2/\kappa$, where $\kappa$ is the string tension $\kappa\sim\mathcal{O}(1)$ GeV/fm.  Ropes can have a higher tension than single strings, which is what gives rise to strangeness enhancement.   This higher tension can also affect other aspects of fragmentation, though as will be true for all the models described in this section, observables that are largely insensitive to hadronization may not be very sensitive to this modification\footnote{In particular, this means that if these models do not predict jet modification, we cannot rule out such effects at the LHC - none of these models currently explains all observed collective effects in $pp$.  However, they serve as useful benchmarks.}.  We use the rope hadronization plugin~\cite{Bierlich:2016faw} to Pythia 8.226~\cite{Sjostrand:2006za,Sjostrand:2014zea} to model collective effects with ropes\footnote{This model does not include shoving~\cite{Bierlich:2016vgw}, which has been shown to qualitatively explain the ridge.  Recent work to incorporate shoving into Pythia 8 is promising and would provide an additionally useful benchmark for jet studies in the future when it is ready~\cite{Bierlich:2017vhg}.}.

While the rope hadronization model is an explicit attempt to describe some collective effects, color reconnection (CR) models that attempt to model $1/N_\text{c}^2$ effects beyond the naive $N_\text{c}=\infty$ limit can also result in collective phenomena.  For example, the default CR model in Pythia 6~\cite{Sjostrand:2006za} has been shown to produce flow-like effects in $pp$ collisions~\cite{Ortiz:2013yxa}.   This has also been demonstrated~\cite{Bierlich:2015rha} for a more recent QCD-inspired model of color reconnection~\cite{Christiansen:2015yqa}.  Both the new and original CR models are studied by either switching to the new model or turning off CR completely. 

A third approach to collective effects draws inspiration from thermodynamics~\cite{Fischer:2016zzs}.  Nearby strings increase an effective temperature that describes the distribution of hadrons and their momentum spectra.   This model also includes rescattering (secondary parton collisions after the first one), which is inspired by the formation of a dense hadronic gas when the string density is sufficiently high.

The above models are simulated by modifying Pythia 8.226~\cite{Sjostrand:2006za,Sjostrand:2014zea} with the default tune~\cite{Skands:2014pea} for $pp$ collisions at\footnote{We focus on 7~TeV since this is the LHC dataset with the lowest pileup, and is more suitable for the type of analysis discussed here.  However, if one restricts to track-only observables, the high pileup data may also be usable by only looking at the tracks associated to a single vertex.} $\sqrt{s} = 7$ TeV.   None of the models were explicitly developed or tested for $pp\rightarrow Z$+jets, but they can still provide useful benchmarks.  Other simulators have been designed to incorporate collective effects, such as Dipsy~\cite{Flensburg:2011kk} and EPOS LHC~\cite{Pierog:2013ria}, but they can only simulate minimum bias events.  The rope hadronization model will be used as the prototypical model of collective effects and the other Pythia modifications will be shown as well.  Appendix~\ref{sec:params} documents all of the parameters used for the various models.

\subsection{Event Reconstruction}
\label{sec:eventreco}

The $Z$-boson is required to decay into muons and $|m_{\mu\mu}-m_Z|<15$ GeV.  Stable particles ($c\tau\leq 10$ mm) excluding muons and neutrinos are clustered into jets with FastJet 3.1.3~\cite{Cacciari:2011ma} using the anti-$k_t$ algorithm~\cite{Cacciari:2008gp} with a jet radius of $R = 0.4$.  Unstable strange hadrons are assigned to jets via ghost association~\cite{Cacciari:2008gn}.  Jet catchment areas are calculated using the median area from the Voronoi method applied to $k_t$ jets clustered from particles out to $|\eta|=2$.  Signal jets are required to have $p_\text{T}>20$ GeV and $|\eta| < 2$. For some technical plots below we shall also use `soft jets', with $10$ GeV $<p_\text{T}<20$ GeV (too low to be reconstructed in practice).  All events are required to have exactly one signal jet and $|\Delta\phi(\text{jet},Z)|>1$ rad to reduce the likely presence of a second jet that is below threshold.  In HI and pPb collisions, the `centrality' of an event is often quantified by the number of particles measured in the event.  Therefore, we study event and jet properties as a function of the measured multiplicity.  There are many ways to quantify the multiplicity:

\begin{enumerate}
\item Total track multiplicity (TTM).  General purpose detectors like ATLAS and CMS have tracking coverage up to $|\eta| < 2.5$ and $p_\text{T}\gtrsim 200$ MeV.  Tracks are excluded if they are within an annulus of $\Delta R < 0.6$ around the signal jet axis. 
\item $Z$-side track multiplicity (ZTM).  Despite an annulus cut around the jet axis, the TTM may be biased by the presence of a jet due to large angle radiation from the parton(s) recoiling from the $Z$.  One way around this is to count the number of tracks that are in the $Z$ boson hemisphere defined by $\cos(\Delta\phi(Z,\text{track})) > 0$.
\item Forward Multiplicity (FM). Even if the tracks from the jet side are removed, the hard $Q^2$ process can still influence the central multiplicity.  Therefore, the number of very forward particles can be used as measure of event activity.  This is a tradeoff between sensitivity to the underlying event activity that might influence the hard $Q^2$ process and a potential bias from the hard $Q^2$ process itself influencing the multiplicity.  We use a cutoff of $4<|\eta| < 5$, which is consistent with the ALICE forward scintillators~\cite{Abelev:2014ffa} and ATLAS/CMS forward calorimeters~\cite{Aad:2008zzm,Chatrchyan:2008aa}.
\end{enumerate}

\begin{figure}[h!]
\centering
\includegraphics[width=0.5\textwidth]{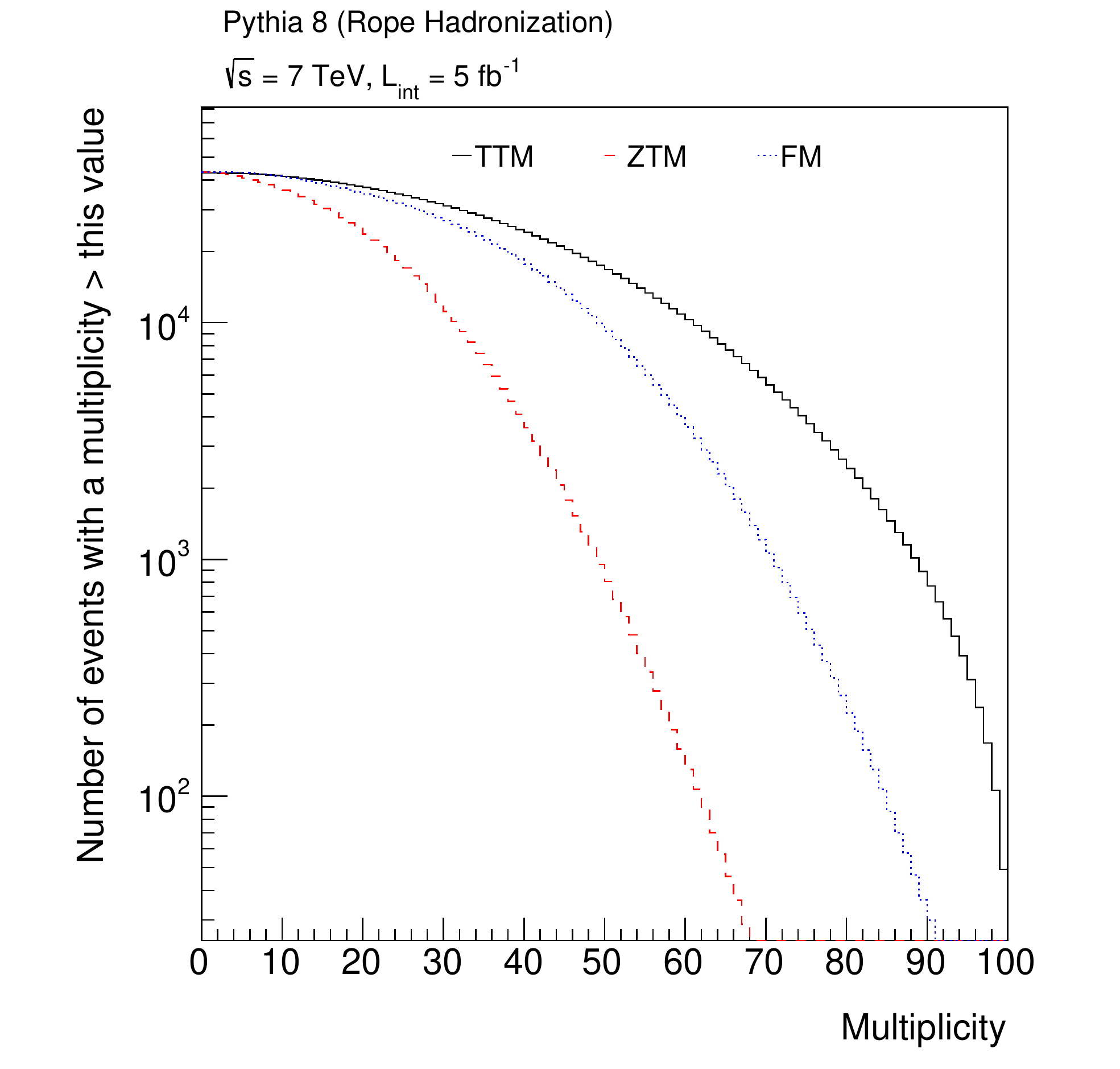}\includegraphics[width=0.5\textwidth]{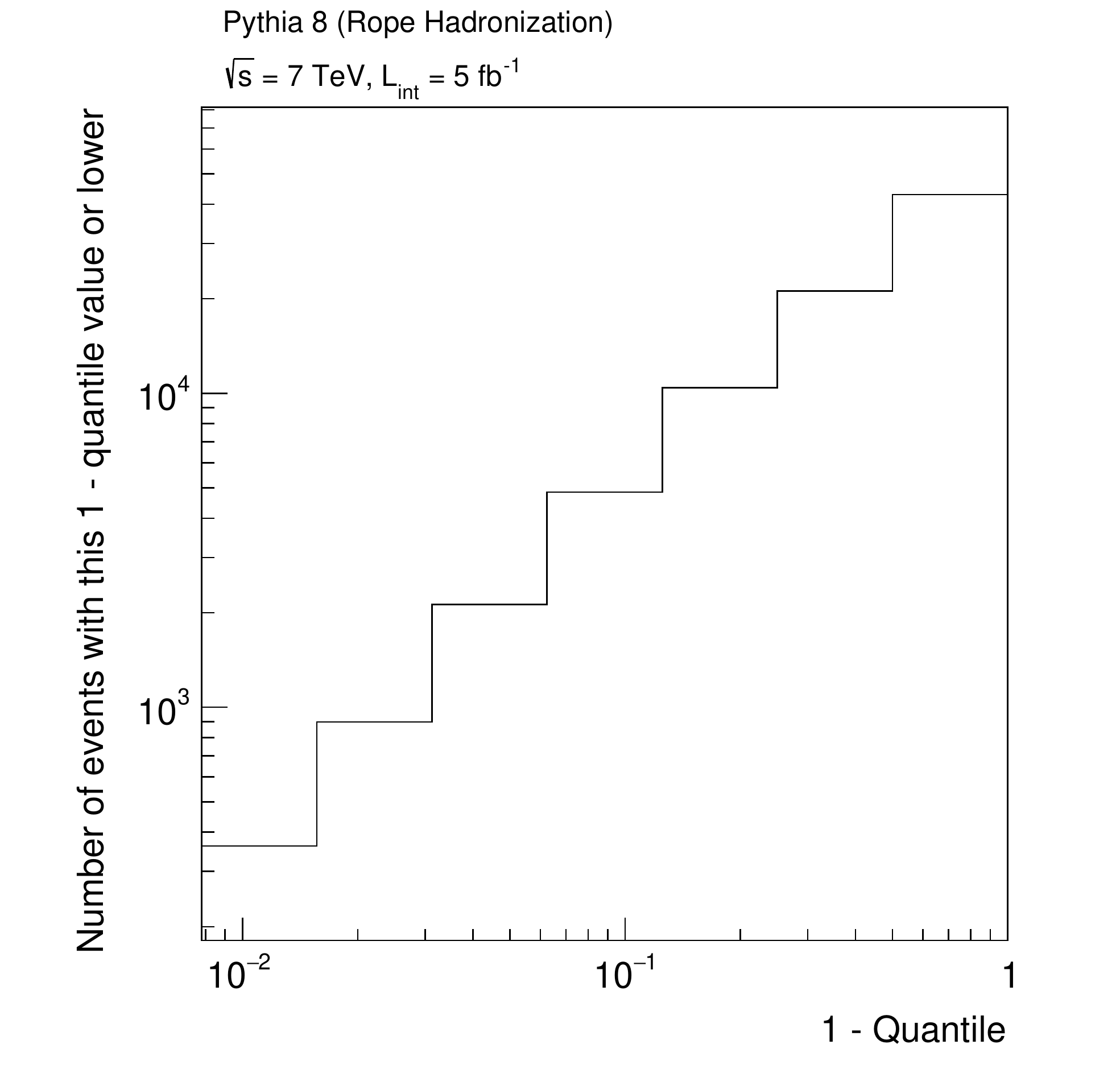}
\caption{Left: The number of expected $Z$ + 1 jet events passing the selection cuts, in 5~fb$^{-1}$, with a given multiplicity threshold defined by TTM, ZTM, or FM.  Right: the number of events that pass a threshold on the multiplicity quantile (the same by construction for all multiplicity types).  Higher multiplicities are on the left. }
\label{fig:fig1}
\end{figure}

Figure~\ref{fig:fig1} shows the number of predicted events with a multiplicity defined by TTM, ZTM, and FM for the Rope hadronization model.  One key advantage of $Z$ (or $\gamma$)+jets in $pp$ collisions versus $pPb$ is that the integrated luminosities collected by ATLAS and CMS of the former are much larger than all four experiments' datasets for the latter.   With only the $\sqrt{s}=7$ TeV dataset, of about 5~fb$^{-1}$, there are many hundreds of $Z\to\mu^+\mu^-$ events with a single jet that are in the $>99\%$ percentile of the multiplicity distribution.   The $n^\text{th}$ quantile is defined such that there are a fraction $n$ of events that have this multiplicity or smaller.  It is a useful notion for normalizing the multiplicity to make direct comparisons between definitions.  Events in the 99\% percentile are such that only 1\% of events have a higher multiplicity. 

\begin{figure}[h!]
\centering
\includegraphics[width=0.5\textwidth]{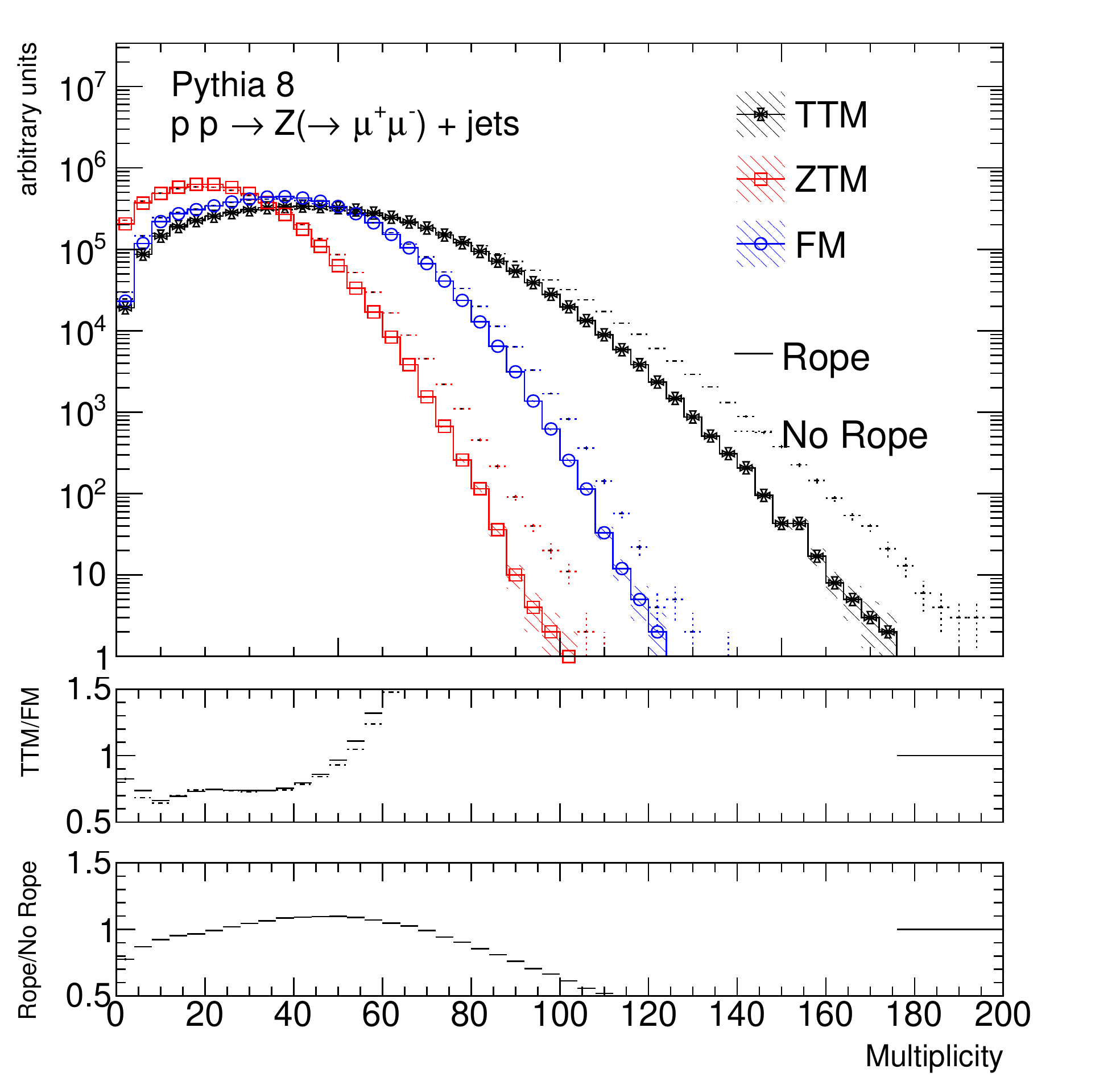}\includegraphics[width=0.5\textwidth]{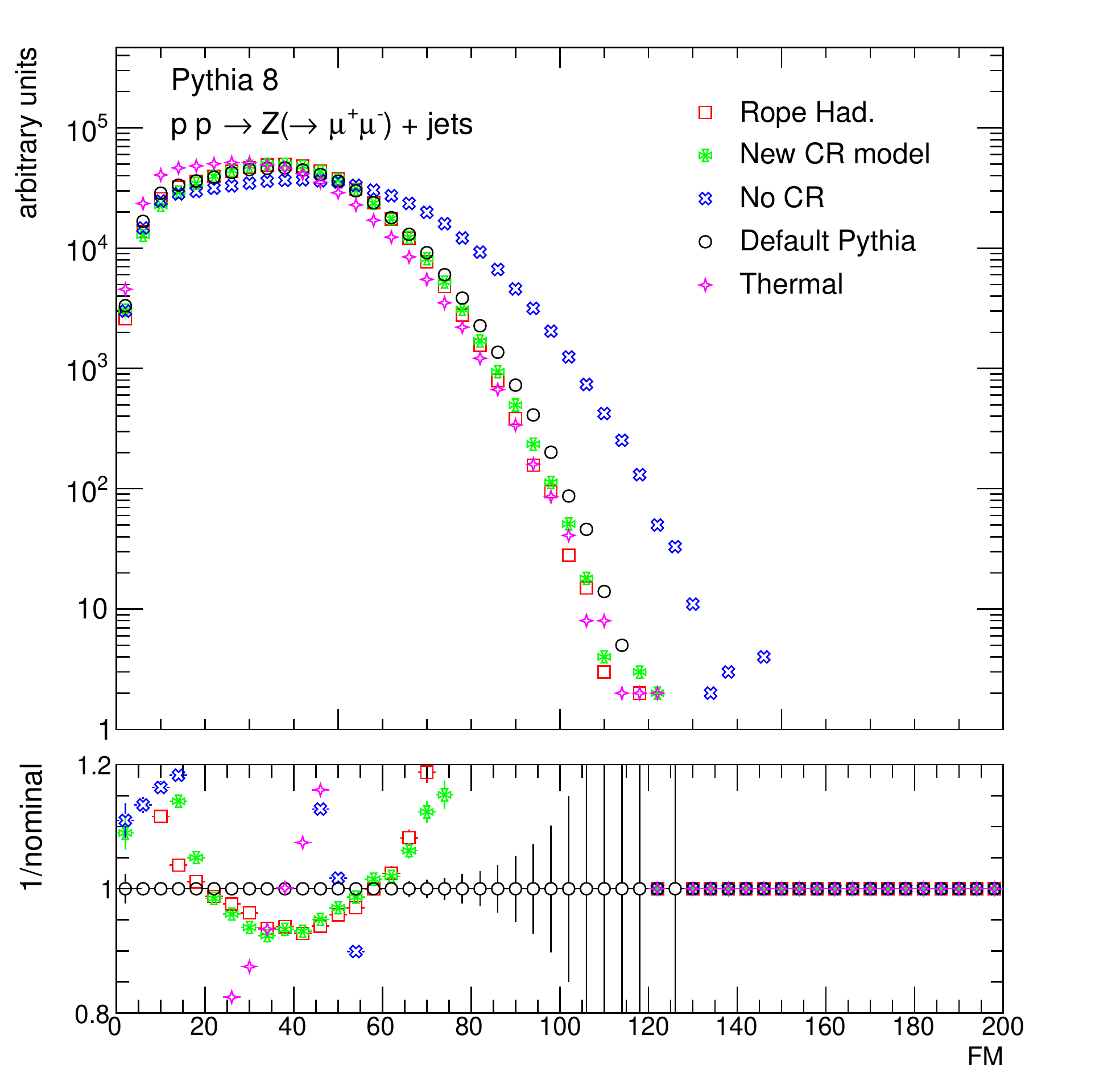} 
\caption{Left: The distribution of the three definitions of multiplicity before for the default and Rope hadronization models.  Right: a comparison of the FM for the various models described in Sec.~\ref{sec:coleffects}.}
\label{fig:multiplicity}
\end{figure}

The actual multiplicity distributions for the three definitions are shown in Fig.~\ref{fig:multiplicity}.   By construction, ZTM is less than or equal to the TTM and is typically a factor of two smaller.  The Rope hadronization model predicts a different multiplicity distribution than the nominal; to control for any effects to these differences when comparing observables in multiplicity ranges, the multiplicity distribution for the  standard hadronization is re-weighted to match the Rope distribution.  The median TTM/ZTM/FM multiplicities are 44/22/36, respectively.   The right plot of Fig.~\ref{fig:multiplicity} shows the FM distribution for all of the models described in Sec.~\ref{sec:coleffects}.  

\clearpage
\section{Results}
\label{sec:results}

\subsection{Strangeness Enhancement}
\label{sec:strangeness}
\begin{figure}[h!]
\centering
\includegraphics[width=0.48\textwidth]{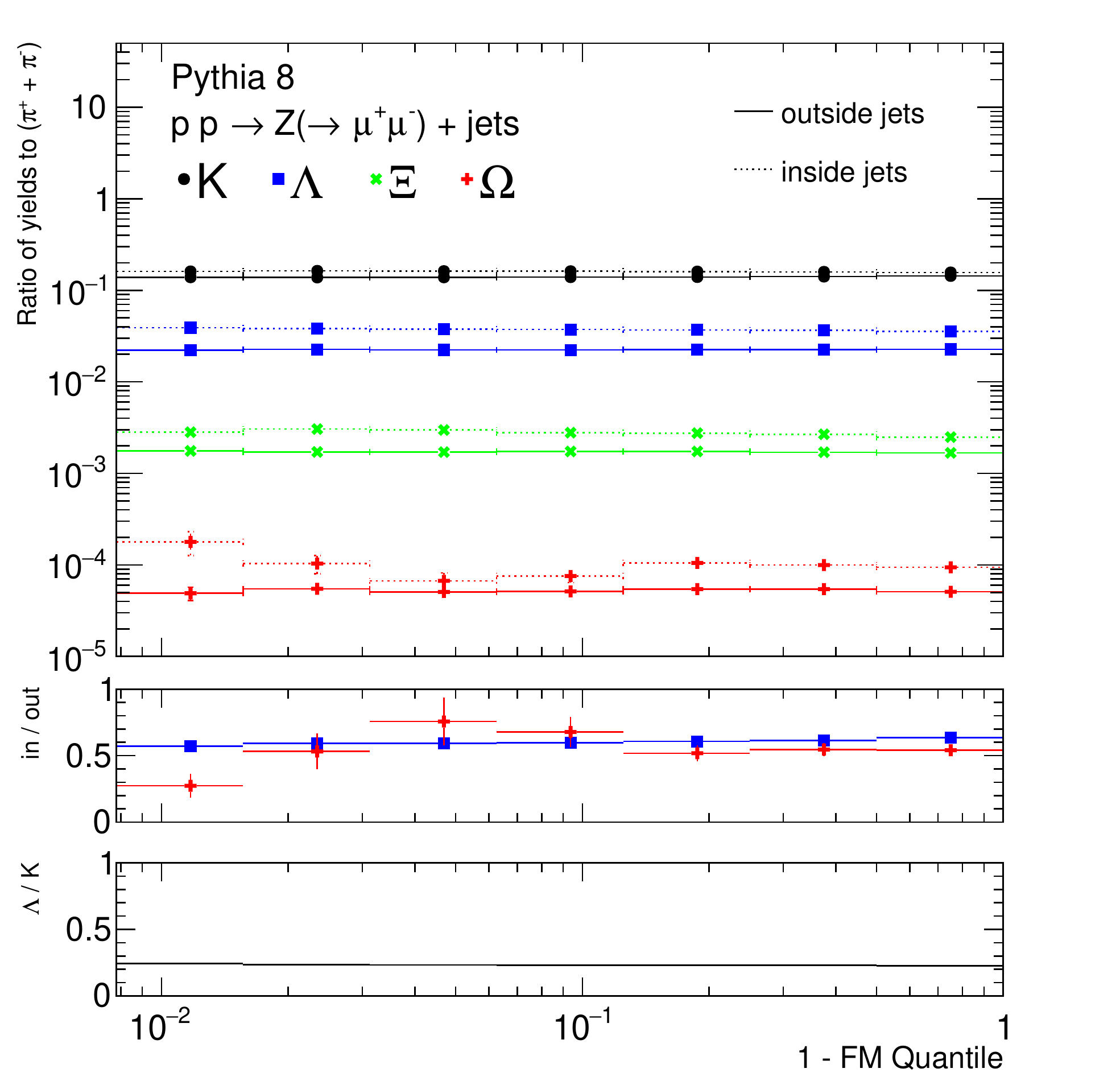}\includegraphics[width=0.48\textwidth]{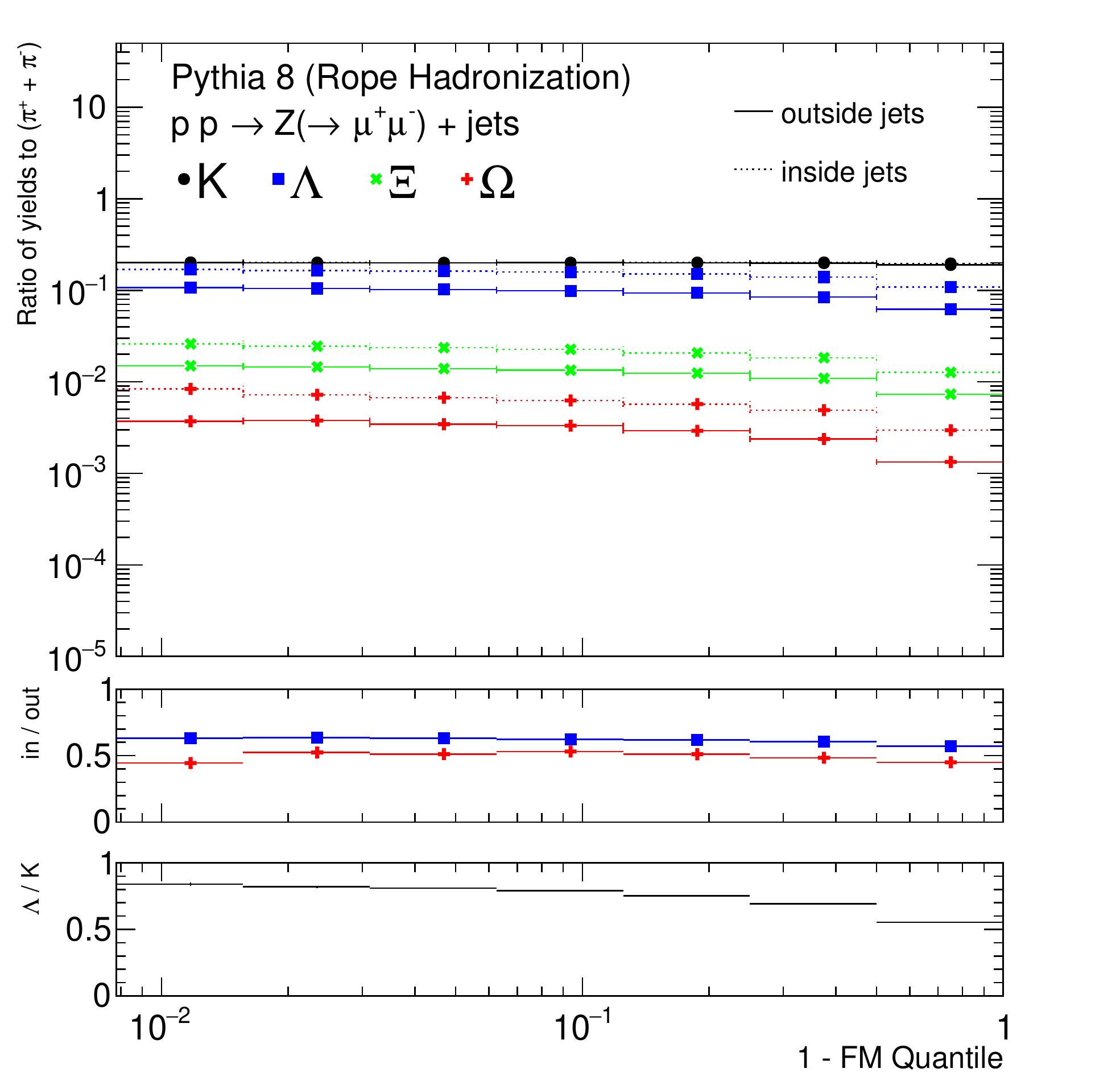}
\caption{The relative abundance of strange hadrons as a function of the FM quantile.  Low values of the $x$-axis correspond to extreme multiplicities; the rightmost bin captures events with multiplicities at or below the median FM.  The left plot shows the trends without the Rope hadronization model while the right plot has the Rope model enabled.}  
\label{fig:strangeTTM}
\end{figure}

We present here some strangeness (and baryon) enhancement variables, considering the multiplicities of strange hadrons produced inside and outside the leading jet.  Figure~\ref{fig:strangeTTM} shows the ratio of various strange hadron yields as a function of the FM criteria (TTM and ZTM are in App.~\ref{sec:altmult}). The left (right) panels represent the case of default (Rope) Pythia hadronization. For all definitions of underlying track multiplicity we notice similar behaviors: no evidence of strangeness enhancement in the case of pure Pythia, compared to the expected clear enhancement in the case of Rope fragmentation. We note the overall increase of strange production in the case of Rope fragmentation with respect to Pythia, independently of the track multiplicity.   

In order to study strangeness and baryon enhancement inside jets, it is useful to subtract the contribution from the underlying event (UE) using an areas-based approach as described in Sec.~\ref{sec:eventreco}.  The following corrected multiplicity and momentum ratios are defined:

\begin{align}
\label{eq:correction}
\frac{n_\text{cor}^{h_1}}{n_\text{cor}^{h_2}}=\frac{n^{h_1}-\rho_\#^{h_1}\times A}{n_\text{cor}^{h_2}-\rho_\#^{h_2}\times A},
\end{align}

\noindent where $h_i$ are hadron species, $\rho_\#$ is the background number density and $A$ is the jet area.  Similarly, the corrected momentum fraction carried by identified hadrons is given by

\begin{align}
z^\text{corr}=\frac{p_\text{T}^{h_1}-\rho^{h_1}\times A}{p_\text{T,jet}-\rho\times A},
\end{align}

\noindent where $\rho^{h_i}$ is the background momentum density for hadron species $h_i$ (and $\rho$ is the standard momentum density).  There is some freedom in defining $\rho_\#$ and $\rho$; we have tested both the standard median approach and one based on the mean where the selected jet is removed.  The jets used are the same as for the standard approach - see Sec.~\ref{sec:eventreco}.  Since the number and momentum carried by identified hadrons can be very small (often zero), the median is not a good estimator of the contribution.  Therefore, we use the mean.

\begin{figure}[h!]
\centering
\includegraphics[width=0.49\textwidth]{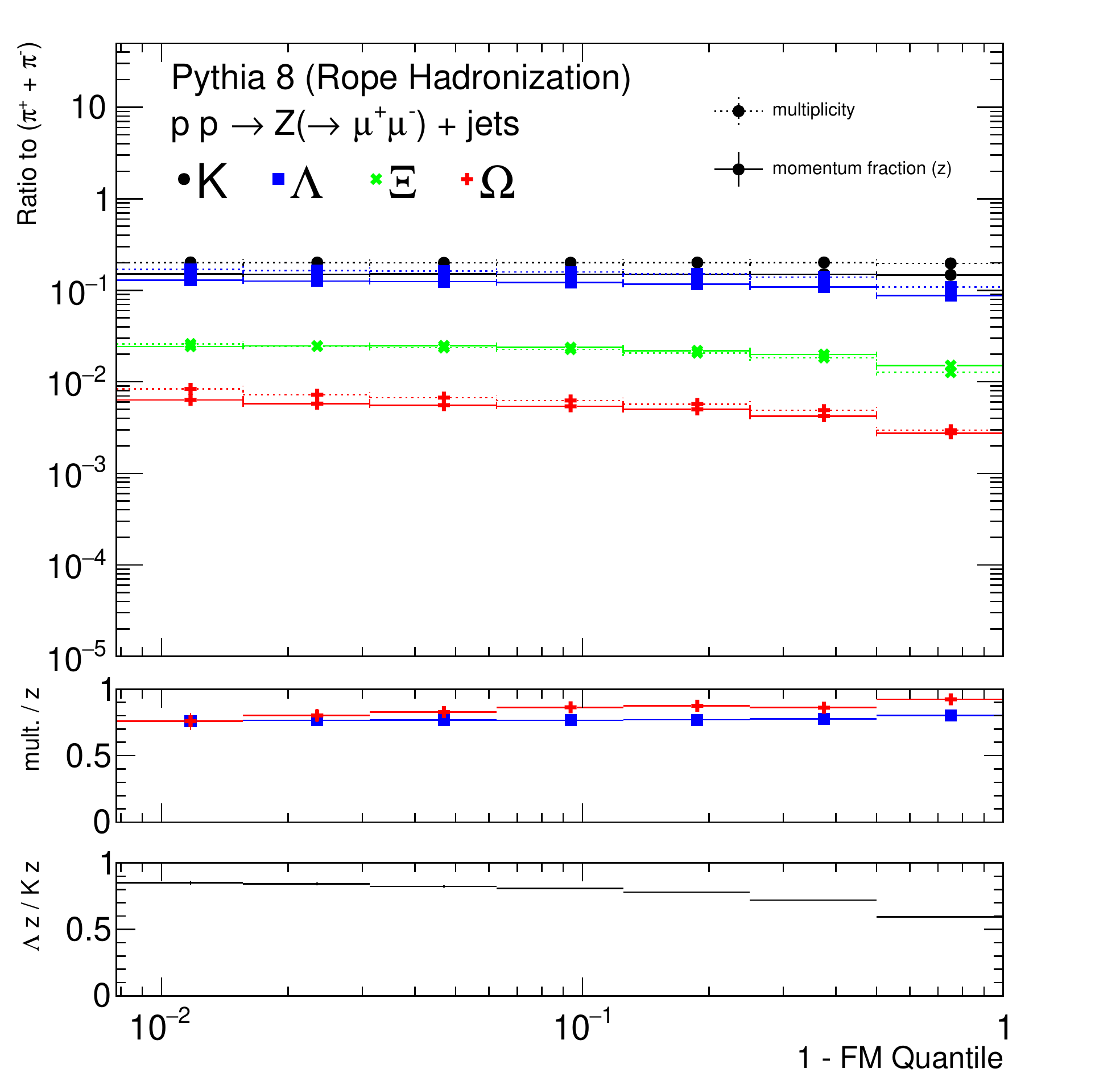}
\caption{Same as Fig.~\ref{fig:strangeTTM}, but comparing the multiplicity inside jets with the average momentum fraction carried by those hadrons.}
\label{fig:strangeAverageFragTTM}
\end{figure}

\begin{figure}[h!]
\centering
\includegraphics[width=0.49\textwidth]{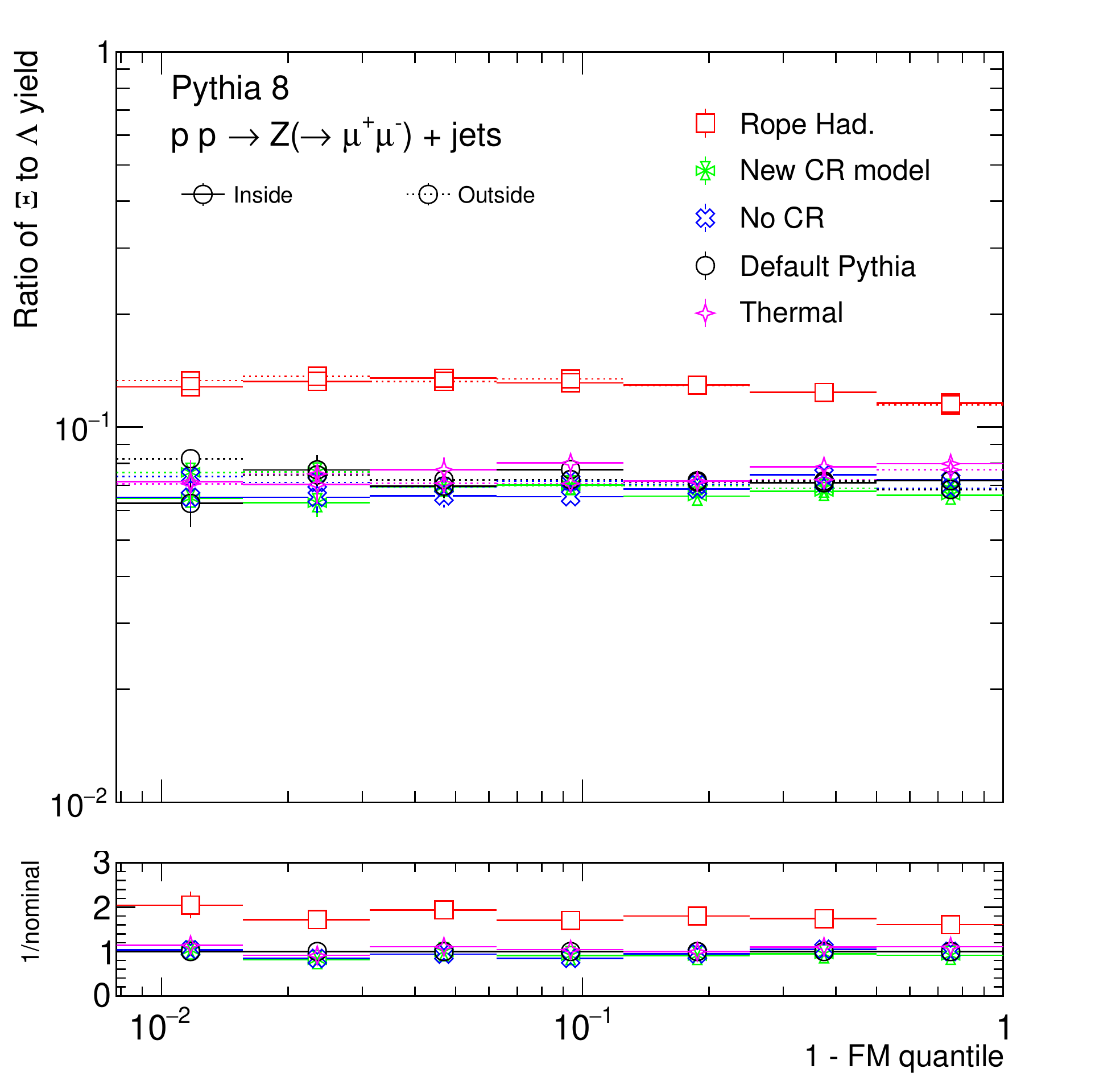}\includegraphics[width=0.49\textwidth]{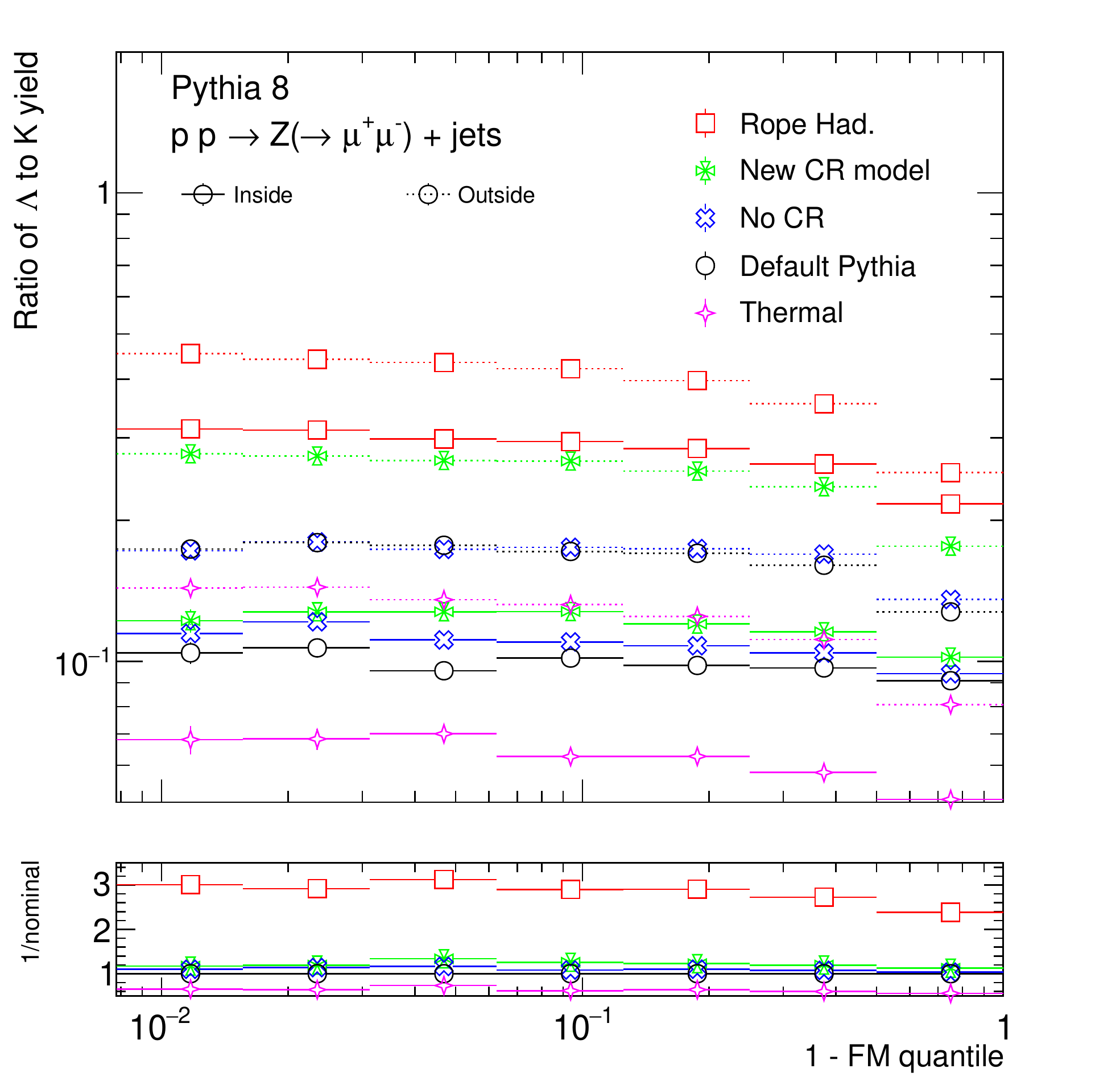}
\caption{Strangeness (left) and baryon (right) enhancement using the $\Xi$-to-$\Lambda$ (left) and $\Lambda$-to-Kaon (right) ratios as a function of the FM quantile for the various models described in Sec.~\ref{sec:coleffects} outside of jets. Low values of the $x$-axis correspond to extreme multiplicities; the rightmost bin captures events with multiplicities at or below the median FM. }
\label{fig:strangemany}
\end{figure}

Figure~\ref{fig:strangeTTM} shows that the ratio of strange hadron fractions inside and outside the jet remains rather constant, for all track multiplicity definitions, for all quantile values, and is also approximately the same for the Rope vs Pythia fragmentations.  These trends are also nearly the same for the multiplicity and the average momentum fraction carried by the hadrons (Fig.~\ref{fig:strangeAverageFragTTM}).   A subset of the comparisons from Fig.~\ref{fig:strangeTTM} focusing on decoupling strangeness enhancement and baryon enhancement are shown in Fig.~\ref{fig:strangemany} for all the models discussed in Sec.~\ref{sec:coleffects}.  The Rope hadronization, thermal hadronization, and new CR models all show signs of strangeness enhancement, with the biggest effect from the rope hadronization.  All of the models predict baryon enhancement at high multiplicity, with the rope and new CR models predicting an increase on top of the default trend.  The thermal model does show an an excess increase over the default model, but the low multiplicity baryon yield is below the default one.  The decomposed ratios suggest that while the strangeness enhancement is predicted to be about the same inside and outside of jets for the Rope hadronization model, the baryon enhancement is much bigger outside of jets\footnote{This may seem to be a contradiction with Fig.~\ref{fig:strangeTTM}, but since the corrections in Eq.~\ref{eq:correction} are applied jet-by-jet, the ratio of the $\Lambda$ to pion line to the $K$ to pion line in Fig.~\ref{fig:strangeTTM} is not equivalent to Fig.~\ref{fig:strangemany}.  It is therefore important to study multiple pairs of ratios to expose the full behavior.}.  Results from ALICE suggest that all of the enhancement inside jets is due to the UE~\cite{Kucera:2015fni}, with no additional modification from the jet itself\footnote{We thank our referee for pointing out that any additional enhancement predicted by our implementation of the Rope model could be a feature of the setup.  Ropes are formed regardless of how fast the strings are moving away from the point of interaction.  This may be inadequate inside jet regions where strings can travel a significant distance prior to hadronizing.}.  The Rope hadronization does indicate additional Baryon enhancement beyond the UE contribution, albeit at a reduced level.  This is also true for  the other models shown in the right plot of Fig.~\ref{fig:strangemany}.

\clearpage

\subsection{Jet Balancing}
\label{sec:jetbalance}
\begin{figure}[h!]
\centering
\includegraphics[width=0.49\textwidth]{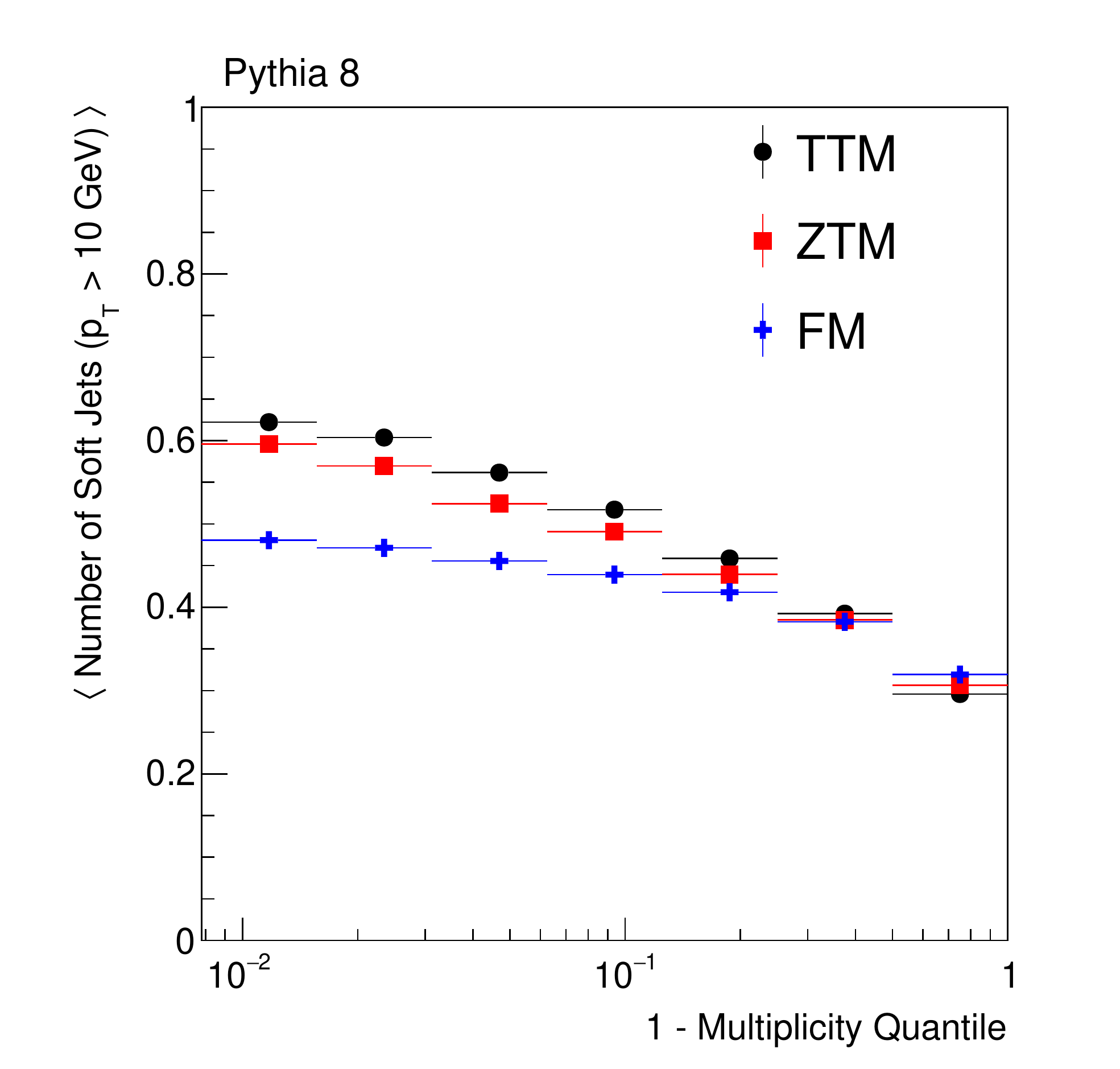}\includegraphics[width=0.49\textwidth]{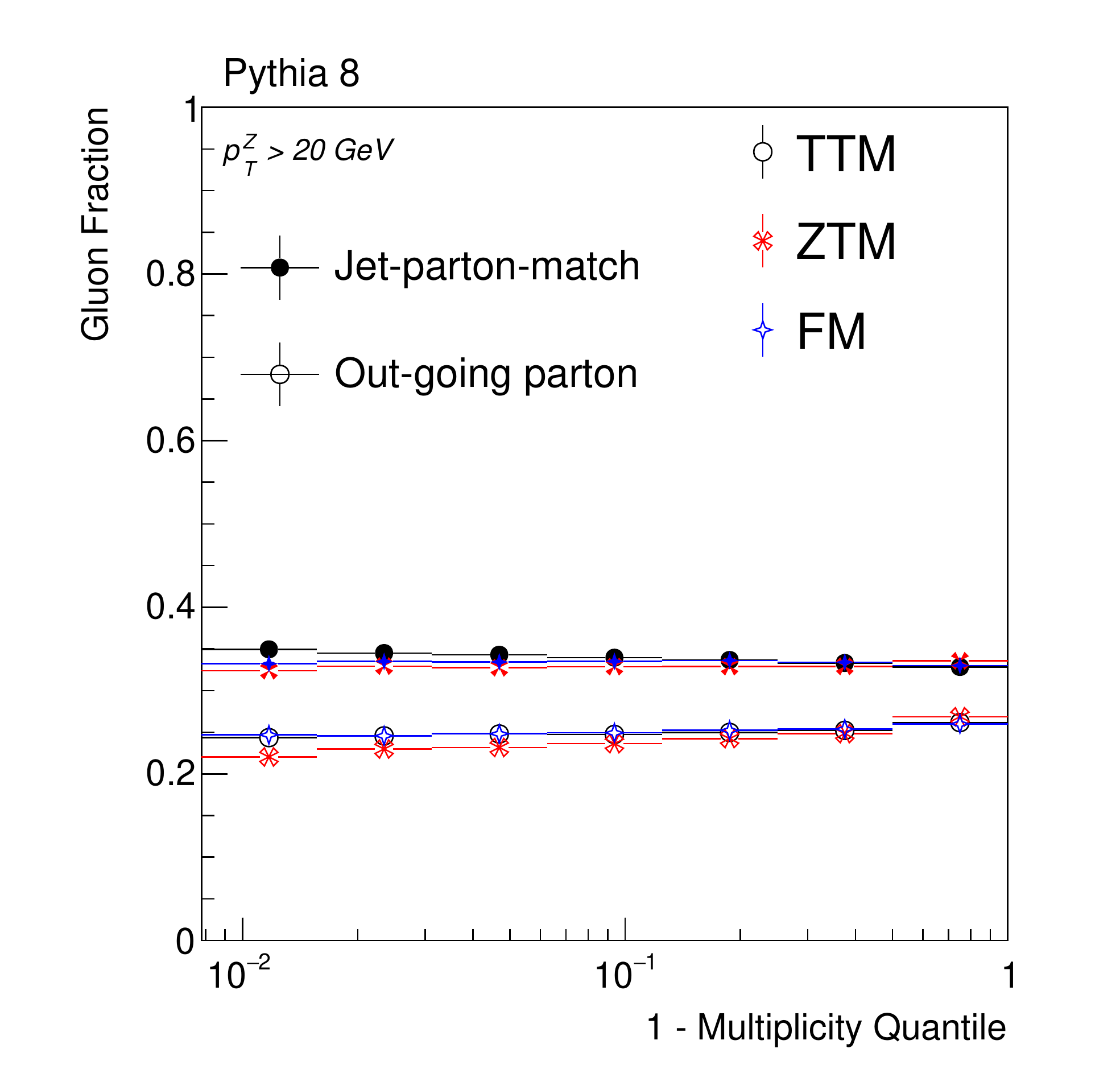}
\caption{Left: The number of soft ($10\,$GeV$<p_\text{T,$J$}<20\,$GeV) jets as a function of the multiplicity quantile for the three multiplicity definitions.  Right: the gluon composition of the hard jet as a function of the multiplicity quantile.  There is no unique way to define the quark or gluon nature of a jet; we show two different definitions: \textit{jet-parton-match} is the type of the highest energy parton found within the $\Delta R < 0.4$ of the jet axis and \textit{out-going parton} is the type of the out-going parton from the matrix element.  Higher multiplicities are to the left.}
\label{fig:radiation}
\end{figure}

\begin{figure}[h!]
\centering
\includegraphics[width=0.49\textwidth]{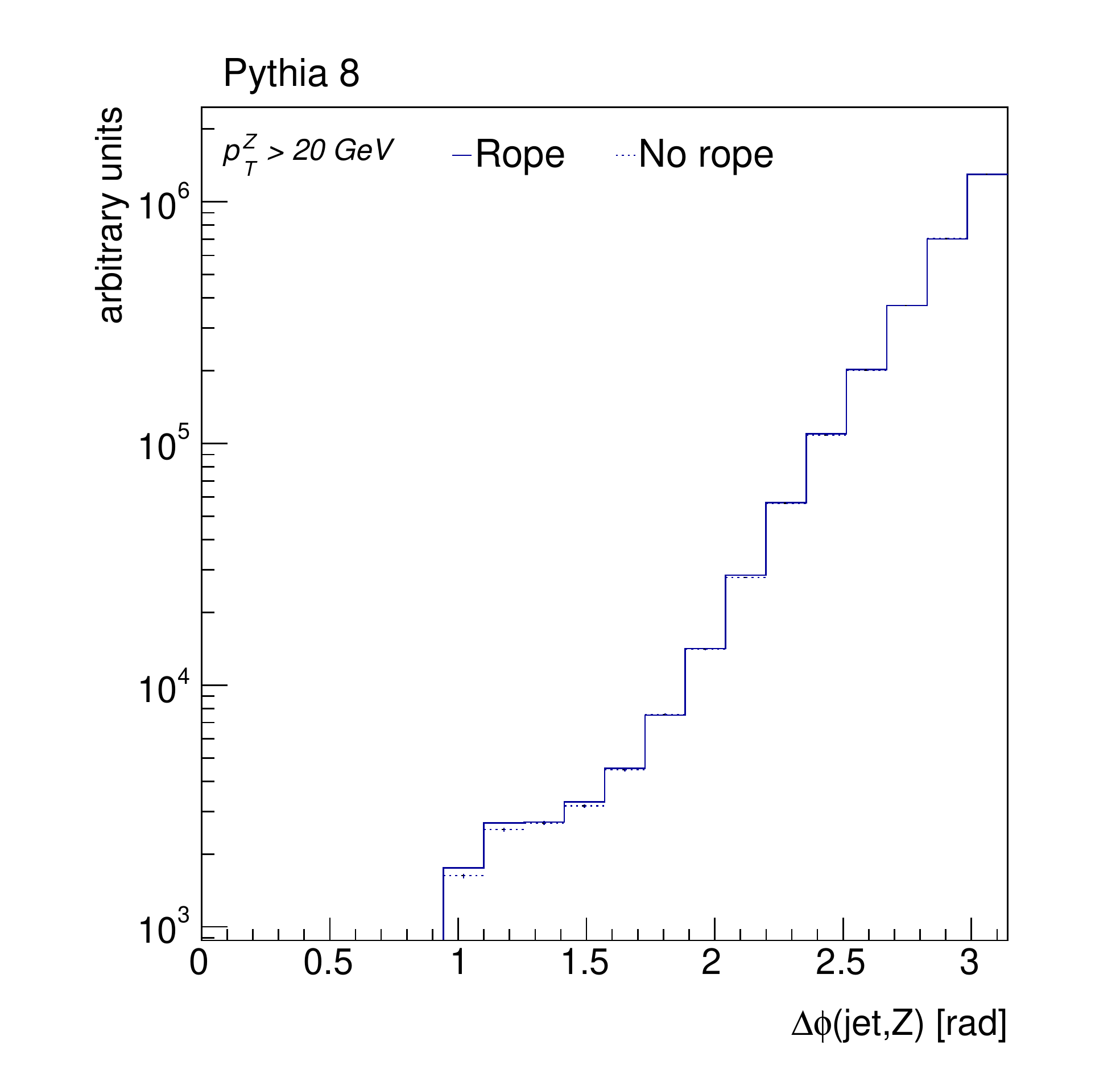}\includegraphics[width=0.49\textwidth]{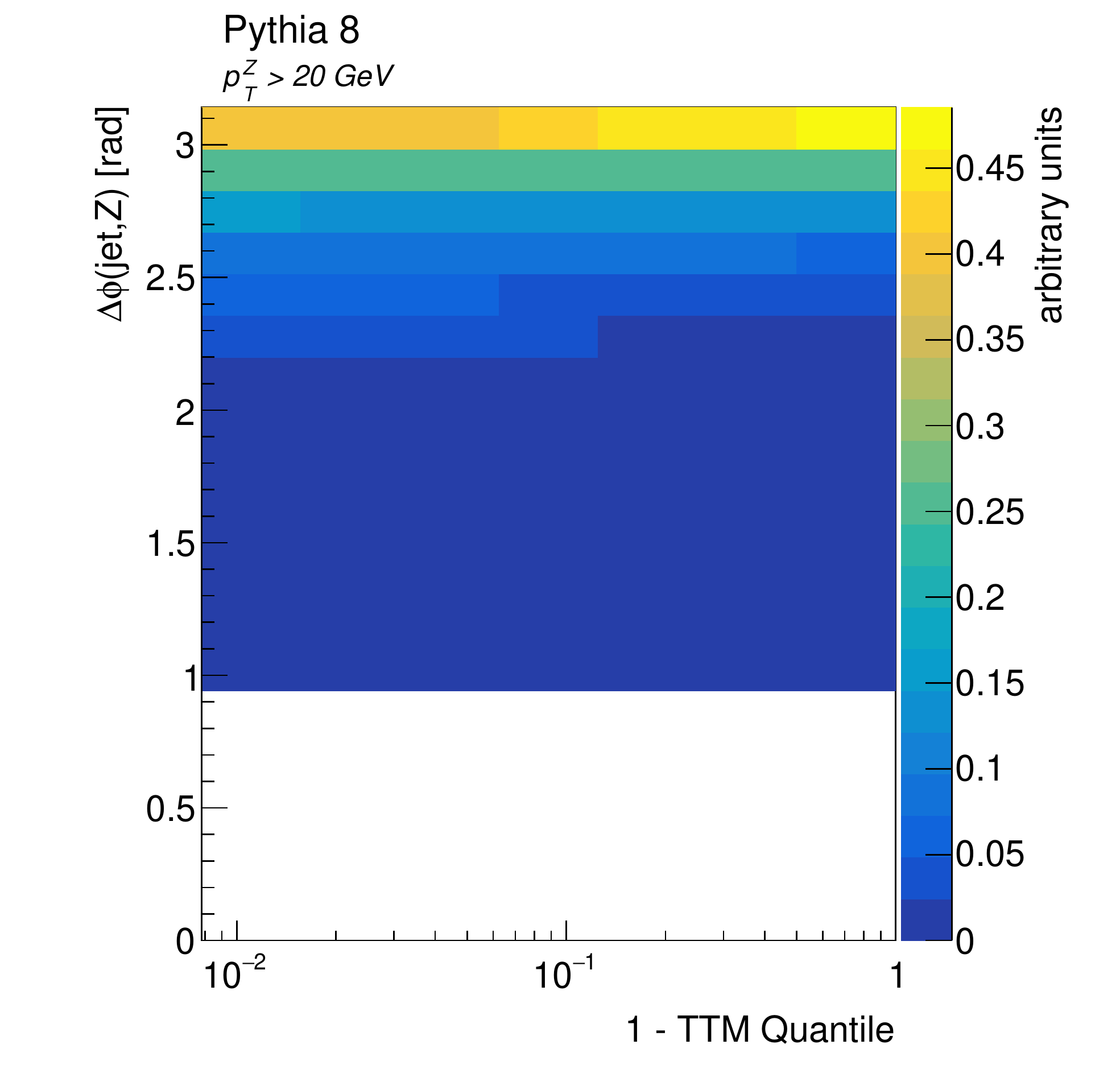}
\caption{Left: The azimuthal angle between the hard jet and the $Z$ boson.  Most events are nearly back-to-back; events are vetoed in the region where there is a strong additional recoil (even if it does not form a jet above threshold). Right: the multiplicity dependence of the azimuthal angle distribution.  There is a slight broadening at higher multiplicities (on the left); this is not present for FM. }
\label{fig:figdeltaphi}
\end{figure}

\begin{figure}[h!]
\centering
\includegraphics[width=0.45\textwidth]{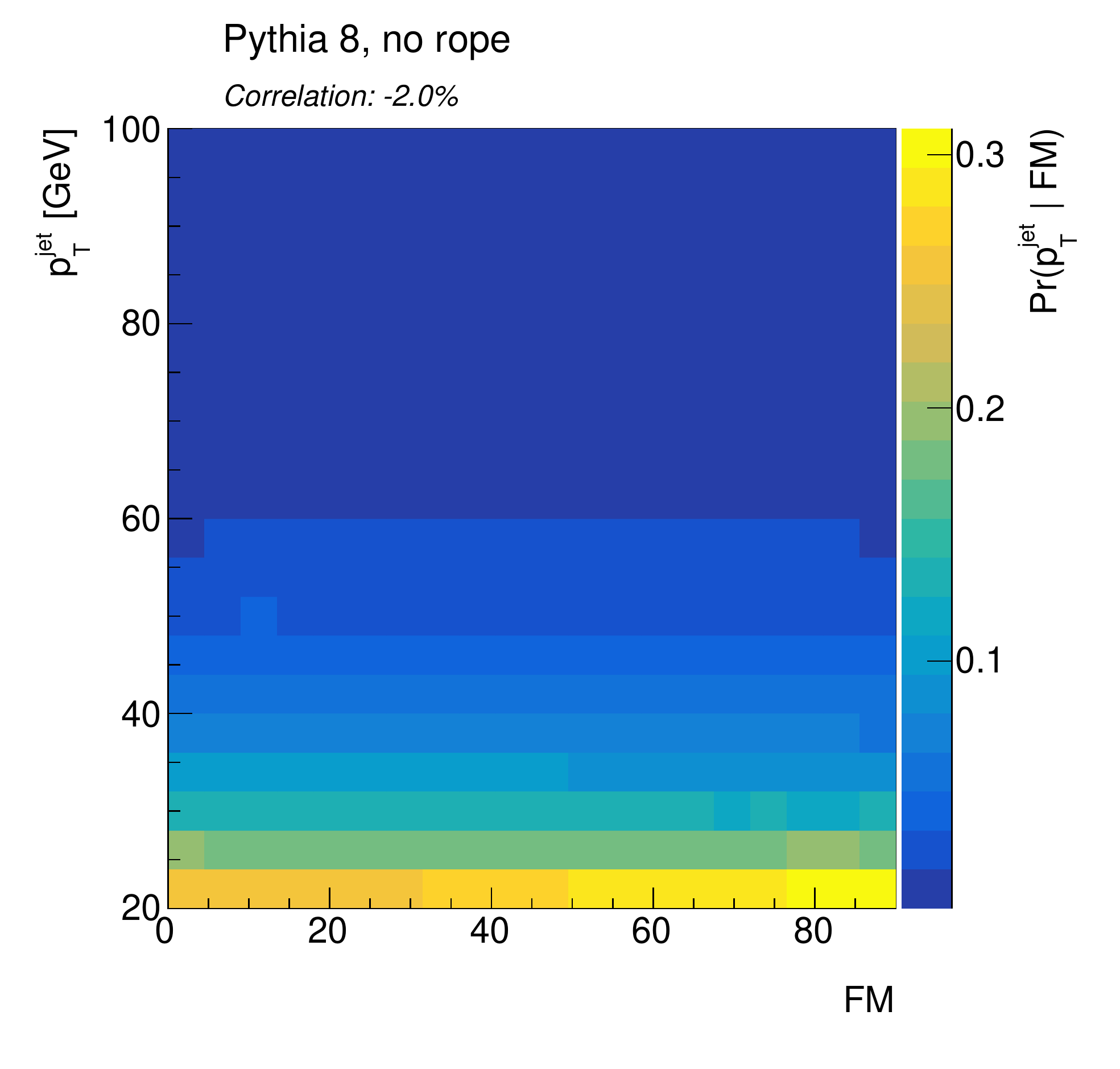}\includegraphics[width=0.45\textwidth]{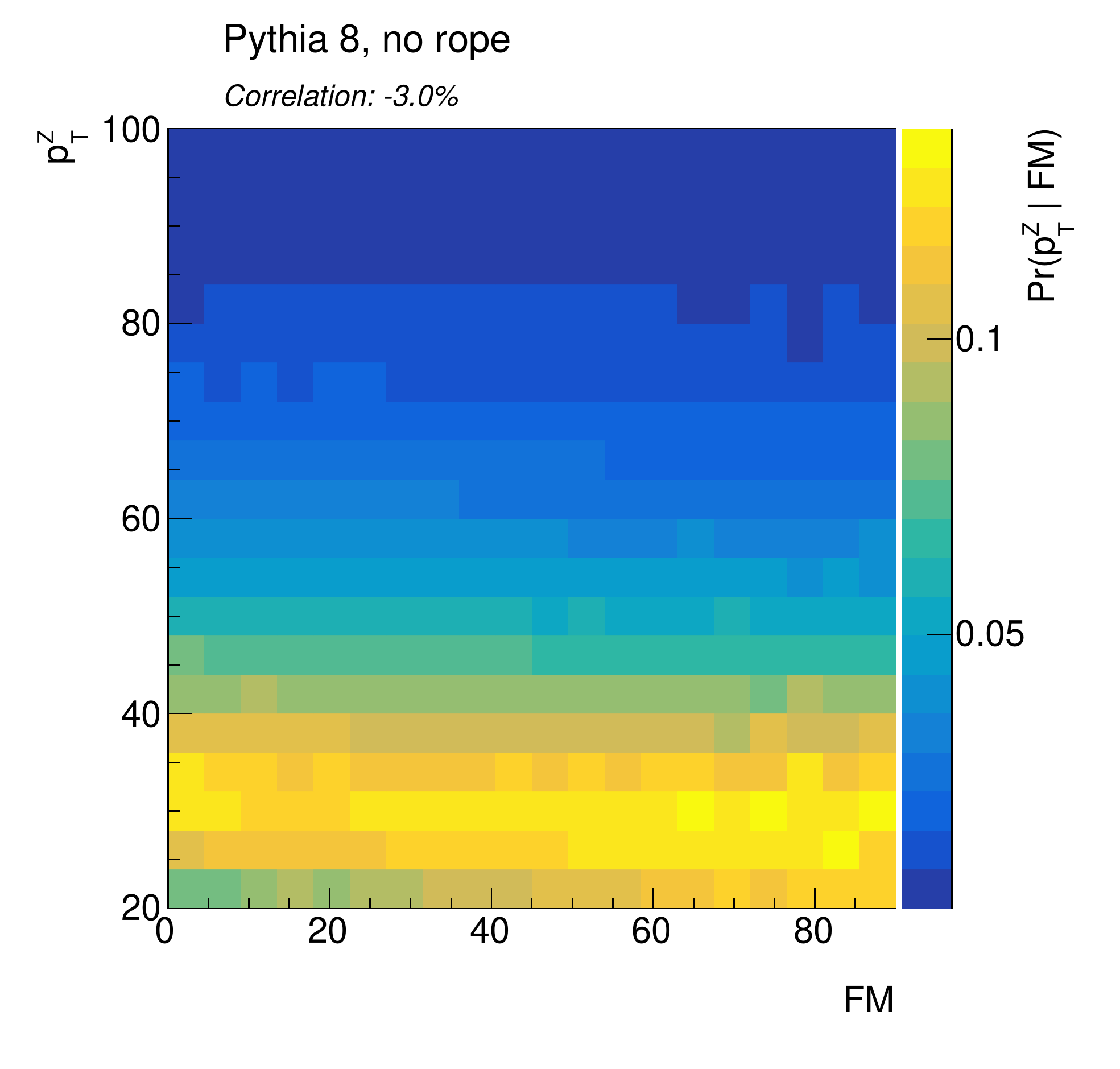}
\caption{The distribution of the jet (left) or $Z$ boson (right) $p_\text{T}$ given the event multiplicity defined by FM.  The linear correlation coefficient is presented at the top of each plot.  The corresponding plots for TTM and ZTM are shown in the appendix, Fig.~\ref{fig:correlationptmultapp}.}
\label{fig:correlationptmult}
\end{figure}

We study in this section a typical observable associated with the presence of a quark-gluon plasma, namely the jet energy loss, leading to an imbalance in the transverse momentum between a jet and its recoil. The cleanest final state in which such phenomenon can be exposed is the recoil of a jet against an electroweak gauge boson, which does not interact with the possible plasma. In particular, we focus on the case of a $Z$ boson decaying to leptons, whose momentum can be well measured, and whose identification is largely free of backgrounds. The study of the $Z$-jet balance at large transverse momentum, as a function of track multiplicity, requires however some caution, since radiation from the hard process will influence the momentum balance, and at the same time it will sculpt the underlying track multiplicity. A different track multiplicity could also reflect a different composition of the initial and final states ($q\bar{q} \to g Z$ vs $qg \to q Z$). All these effects might in principle induce an imbalance that emulates a quenching trend at the highest track multiplicities. The extent of such correlations between track multiplicity, hard radiation and initial state composition is shown in Fig.~\ref{fig:radiation}. The left plot shows the average multiplicity of soft ($p_\text{T,$J$}>10\,$GeV) jets versus the multiplicity quantile, showing that high multiplicity events (lower quantiles) are correlated to a larger radiation activity.  As expected, the correlation is strongest for TTM, then ZTM, and weakest for FM.  This will influence the ratio of the $Z$ boson $p_\text{T}$ to the jet $p_\text{T}$ and can also be observed to broaden the distribution of the angular separation, as in Fig.~\ref{fig:figdeltaphi}.  The right plot of Fig.~\ref{fig:radiation} shows on the other hand a minor, if any, dependence of the gluon final-state fraction versus track multiplicity.  Similarly, there is little correlation between the jet or $Z$ $p_\text{T}$ itself on the event multiplicity, as shown in Fig.~\ref{fig:correlationptmult}.

\begin{figure}[h!]
\centering
\includegraphics[width=0.48\textwidth]{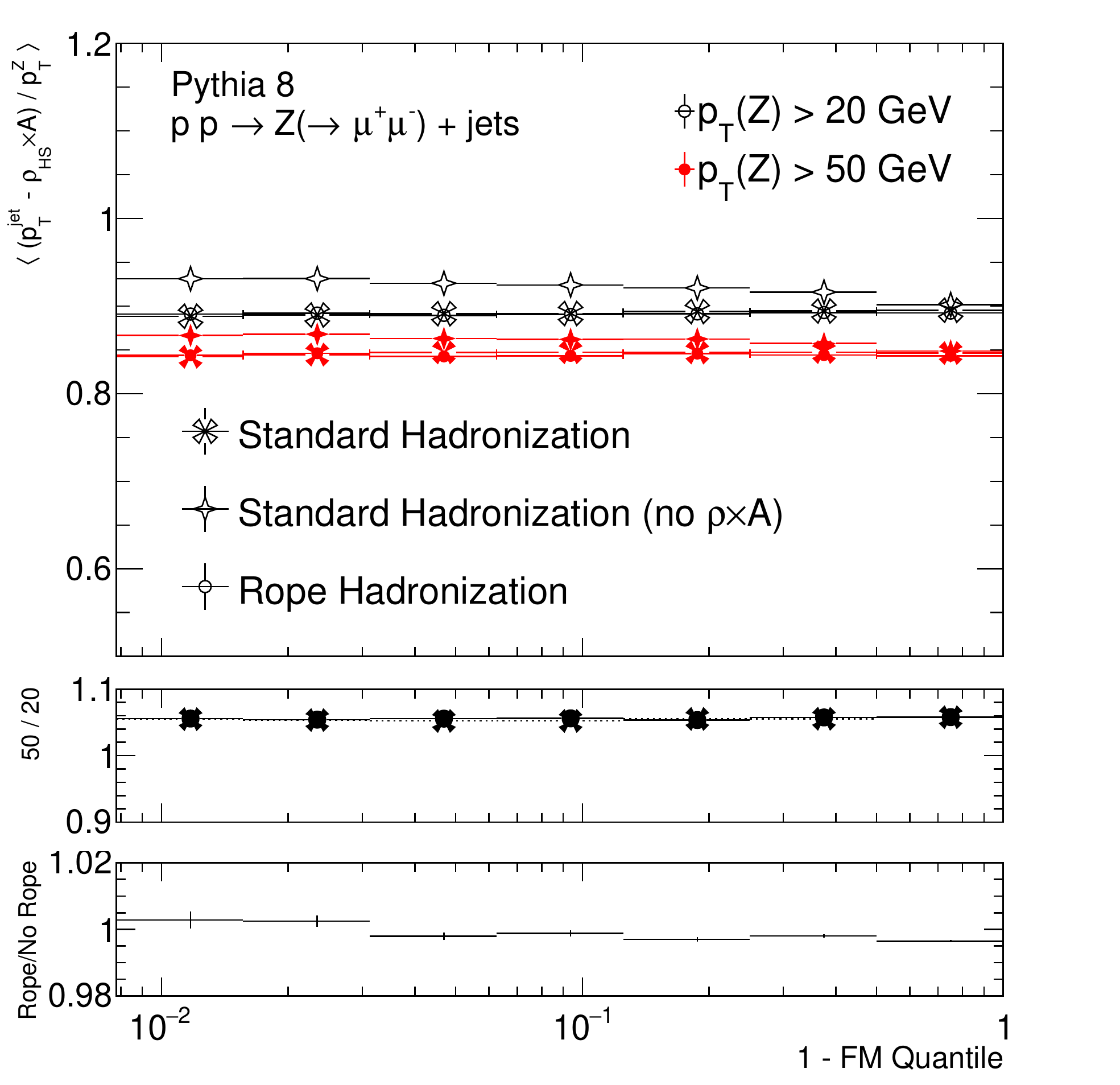}\includegraphics[width=0.48\textwidth]{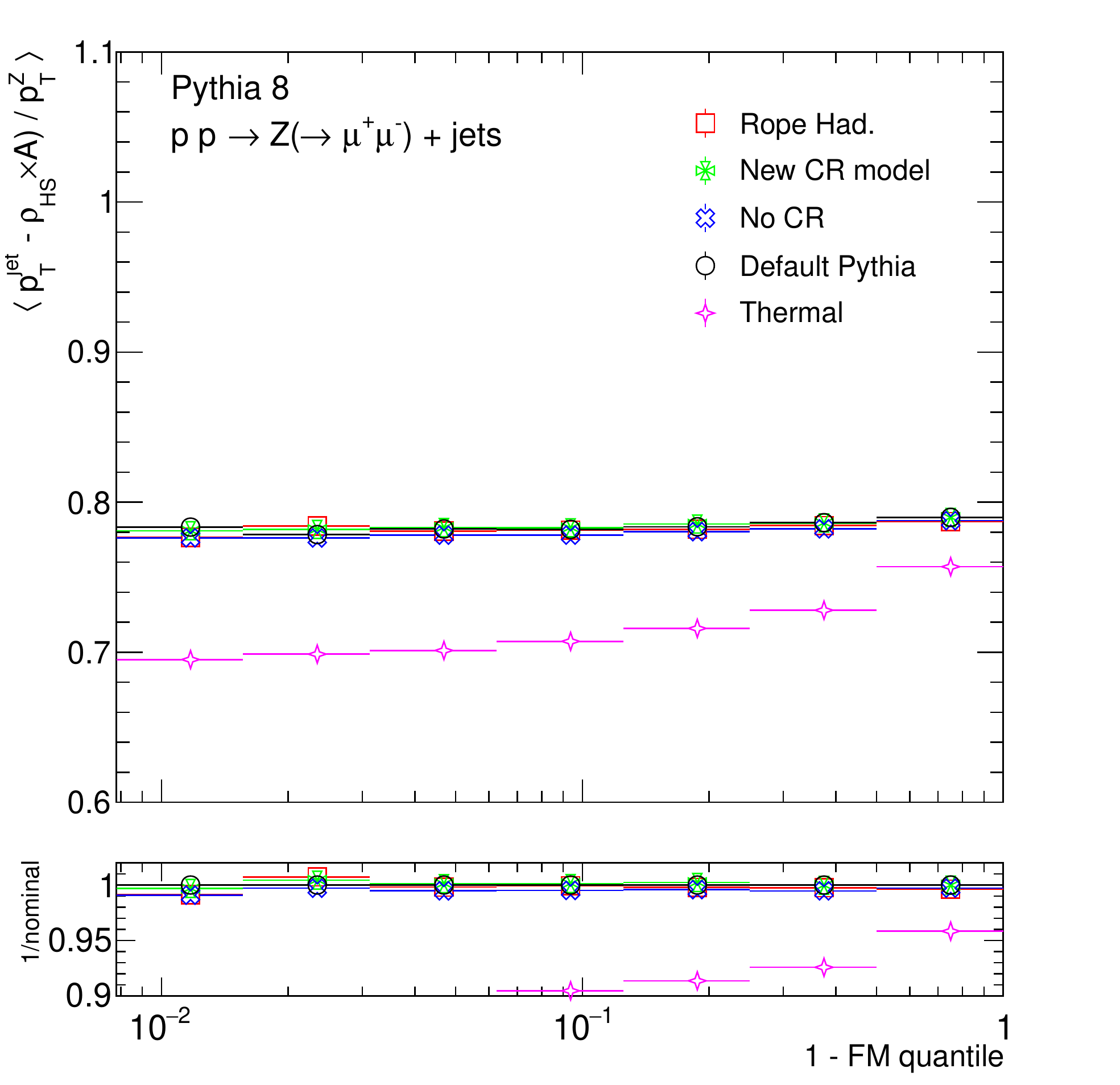}
\caption{The average fractional transverse momentum imbalance between leading jet and $Z$ boson, $x_{ZJ}=p_\text{T,$J$} / p_\text{T,$Z$}$,  as a function of track multiplicity quantile, FM.  Three curves in the upper panels indicate the values for the standard hadronization with and without an area correction as well as the rope hadronization with the correction.  Higher multiplicities are to the left.  The middle panel shows the ratio of the higher to lower $Z$ boson $p_\text{T}$ cut (standard hadronization is shown with a dashed line, rope with solid), both including the areas correction.  Finally, the lower panel shows the ratio between the rope and standard hadronization models with the $p_\text{T,$Z$}>20$ GeV requirement and both with the areas correction.  The corresponding plots for TTM and ZTM are in the appendix, Fig.~\ref{fig:jetsapp}.  The right plot shows similar information for the single $p_\text{T}(Z)>20$ GeV threshold and multiple models.}
\label{fig:jets}
\end{figure}

With these observations in mind, we show in
Fig.~\ref{fig:jets} the average fractional transverse momentum imbalance between leading jet and $Z$ boson, $x_{ZJ}=p_\text{T,$J$} / p_\text{T,$Z$}$,  as a function of track multiplicity quantile, focusing on the least biased multiplicity indicator FM. Results are shown for two thresholds of $p_\text{T,$Z$}>20$ and $>50\,$GeV, considering the cases of Pythia and Rope fragmentation. Except where indicated by the caption ``no $\rho\times$A", an average subtraction of the underlying event activity inside the jet cone is performed.   The distributions with respect to the FM quantile are independent of multipiclity, and appear not to be influenced by a possible hard radiation bias that exists for the TTM and ZTM. This suggests that FM would be a robust variable to explore the possible presence of quenching-induced imbalance.  The right plot of Fig.~\ref{fig:jets} shows that nearly all of the models predict the same trend with multiplicity except the thermal model, which exhibits extreme `quenching'-like behavior at high multiplicity.  


\begin{figure}[h!]
\centering
\includegraphics[width=0.49\textwidth]{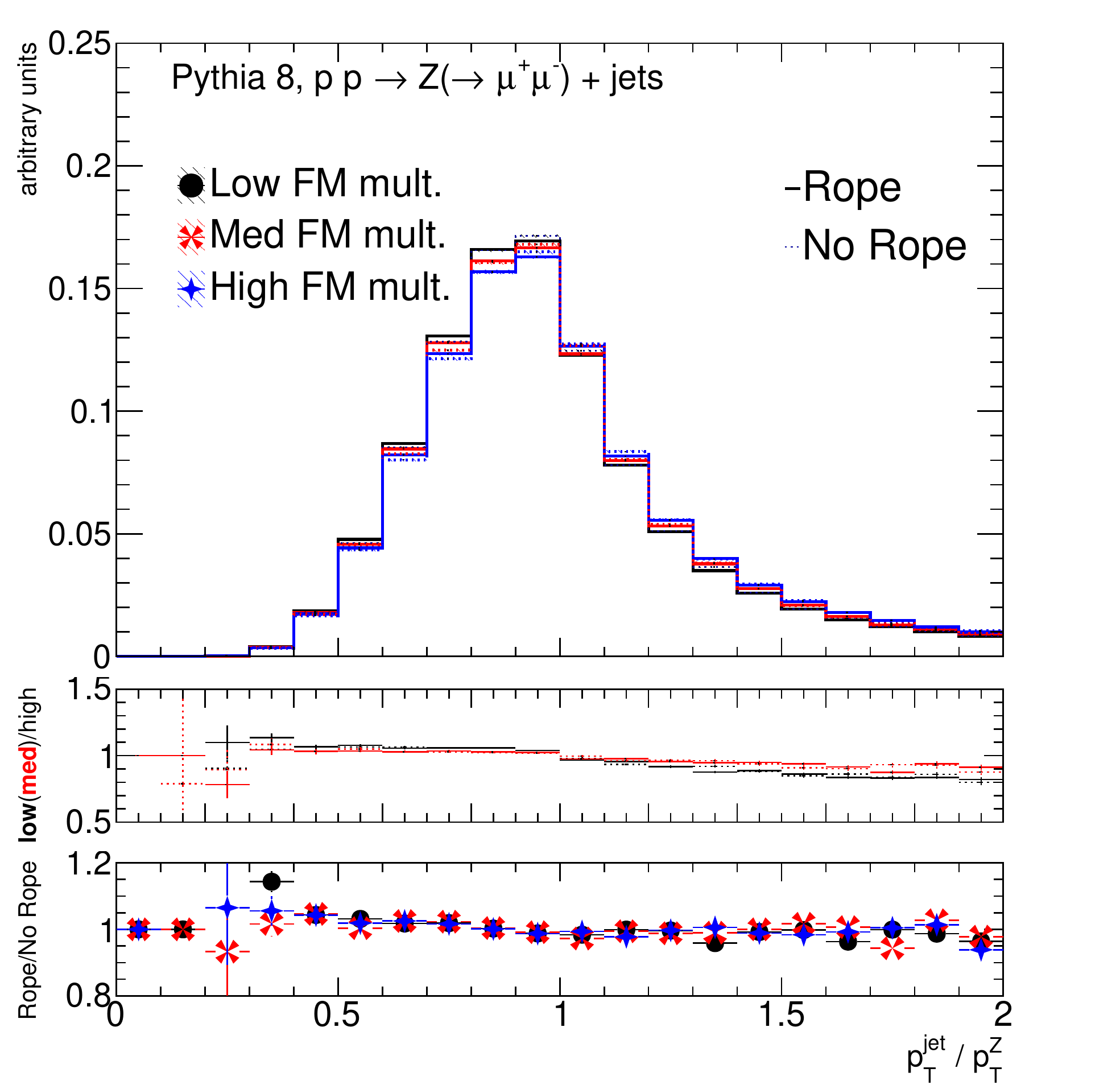}
\caption{The distribution of the ratio $x_{ZJ}=p_\text{T,$J$} / p_\text{T,$Z$}$ for three bins of the TTM (top left), ZTM (top right), or FM (bottom) multiplicity: the 50th, 75th, and 90th percentile.  The middle panel shows the ratio of the distribution for high to low multiplicity for both hadronization models (standard hadronization with a dotted line).  In the lower panel, the ratio between the Rope and standard hadronization models is displayed for all three multiplicity regions.  The corresponding plots for TTM and ZTM are in the appendix, Fig.~\ref{fig:ptratioapp}.}
\label{fig:ptratio}
\end{figure}

Figure~\ref{fig:ptratio} shows the full distribution of the ratio in three bins of multipicity.  Nearly independent of the percentile, the standard deviation of the ratio distribution is about 20\%.  With about 500 jets in the 1\% percentile category (Fig.~\ref{fig:fig1}), the statistical precision in the determination of $x_{ZJ}$ will be $\lesssim1\%$.  The experimental resolution should be comparably small.  The high multiplicity single $pp$ collisions have comparable multiplicity to low/moderate pileup bunch crossings, similar to the levels with early Run 2.  Therefore, one can estimate the uncertainty in the reconstructed jet energy due to pileup as an estimate of the uncertainty for high multiplicity single $pp$ collisions~\cite{Aaboud:2017jcu,CMS-DP-2016-020}. As a figure of merit, a 1~GeV energy loss would correspond, for a 20~GeV jet, to a 5\% effect on $\langle x_{ZJ}\rangle $. 

\clearpage

\subsection{Jet Substructure}
\label{sec:jetstructure}

\begin{figure}[h!]
\centering
\includegraphics[width=0.4\textwidth]{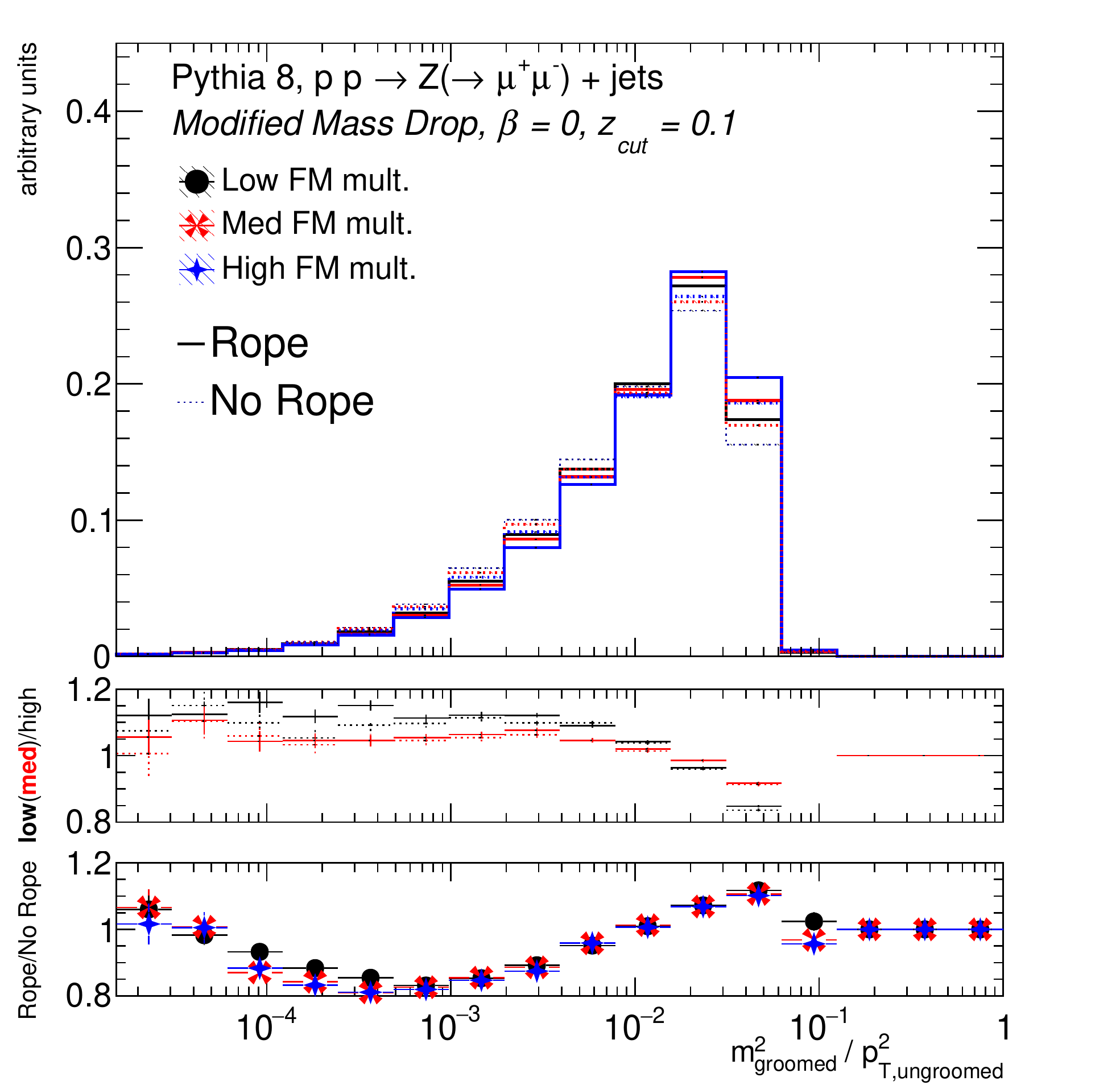}
\includegraphics[width=0.4\textwidth]{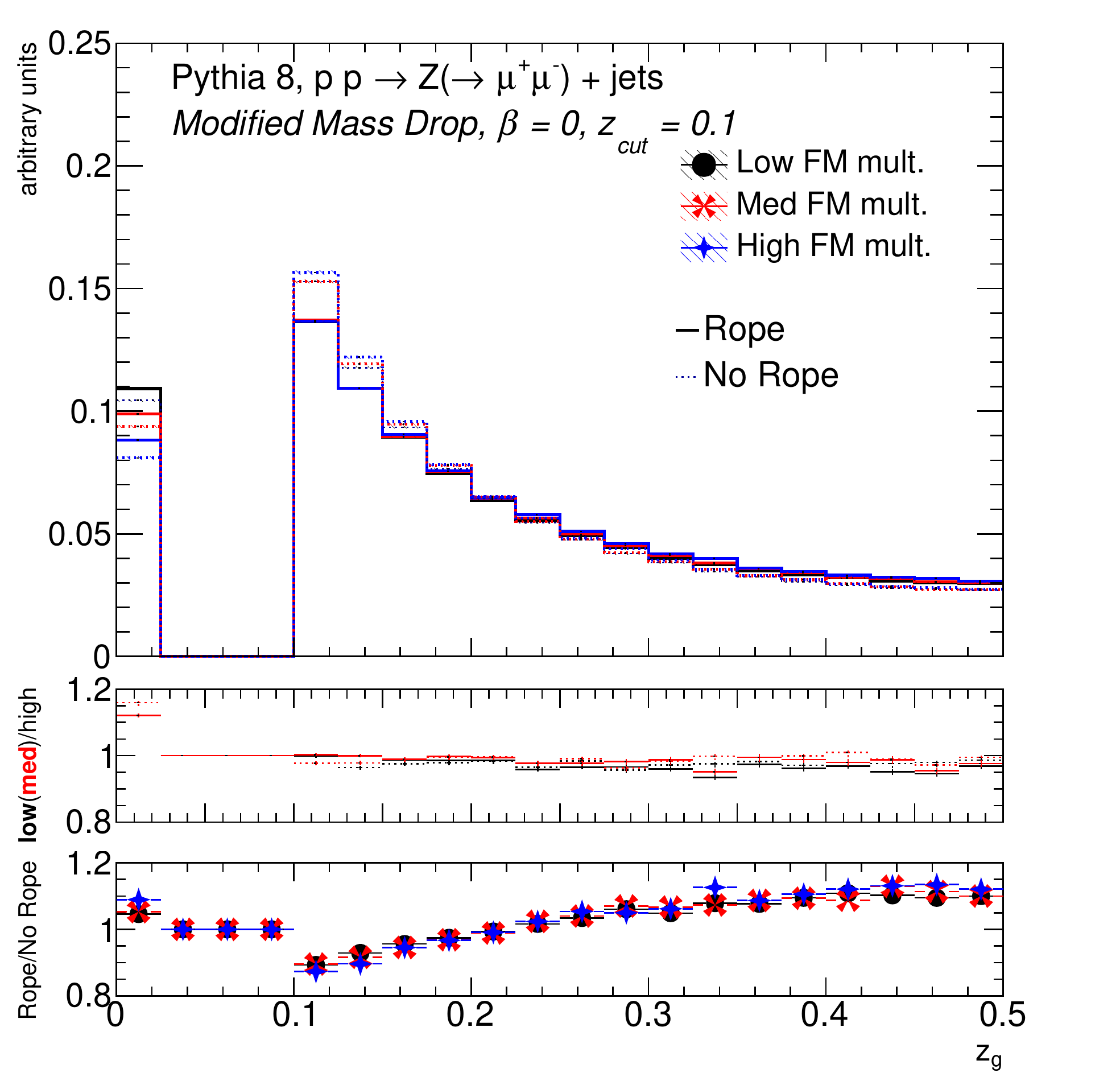}
\caption{Modified Mass Drop (also Soft Drop with $\beta=0$) jet mass (left) and momentum sharing $z_g$ (right) for three bins in FM that correspond to the 50th, 75, and 90th percentiles.  The middle panel shows the ratio of the distributions for high to low multiplicity for both hadronization models (standard hadronization with a dotted line).  In the lower panel, the ratio between the Rope and standard hadronization models is displayed for all three multiplicity regions.  Due to the algorithm value $z_\text{cut}=0.1$, $z_g \geq 0.1$; when the entire jet is groomed away, $z_g=0$.  The corresponding plots for TTM and ZTM are in the appendix, Fig.~\ref{fig:softdropapp}.}
\label{fig:softdrop}
\end{figure}

In addition to reducing the total energy inside a jet, interactions with the QGP in HI collisions distort the radiation pattern.  Scattering with the medium results in jets with a broader distribution of energy and therefore jet substructure tools may be used to search for a QGP in central $pp$ collisions.  The soft drop jet grooming procedure~\cite{Larkoski:2014wba} (the generalization of modified mass drop~\cite{Dasgupta:2013ihk} when $\beta\neq 0$) has gained a lot of recent attention theoretically and experimentally because of its insensitivity to non-global logarithms and robustness to wide angle and soft radiation.  Therefore, soft drop jet observables are useful for probing if the structure of a jet has changed in events with high multiplicity.  In addition to the jet mass, another important soft drop jet observables is the fraction of the groomed jet's momentum carried by the subleading subjet, $z_g$.  This observable has the interesting property that it is independent of $\alpha_s$ at leading order and is directly related to the QCD splitting functions in vacuum~\cite{Larkoski:2015lea}.  Therefore, any change in the $z_g$ distribution in high multiplicity events may be an indication of a modification of the partonic fragmentation function.  Preliminary results from CMS are suggestive of medium-induced modifications of the $z_g$ distribution~\cite{CMS:2016jys}, which are also predicted by various models of jet quenching~\cite{Chien:2016led,Mehtar-Tani:2016aco,KunnawalkamElayavalli:2017hxo,Chang:2017gkt,Milhano:2017nzm}, although there may be\footnote{We say `may be' because the analyses are not identical so differences may arise from the approach.} a tension with preliminary STAR results~\cite{Kauder:2017cvz}.  Figure~\ref{fig:softdrop} shows the distribution of the soft drop mass and $z_g$ as a function of FM with the standard and Rope hadronization models using $z_\text{cut}=0.1$ and $\beta=0$.  The mass does show a dependence on the multiplicity, which may be due in part to a residual contribution within the catchment area of the subjets not removed from grooming.  There is also a significant difference in the shape of the mass distribution between the two hadronization models, though there is little dependence of the difference on multiplicity.   The multiplicity dependence is much reduced in the case of the $z_g$ spectra, and in particular the FM dependence is particularly flat (TTM and ZTM are in the appendix), confirming the smaller radiation bias of the FM distributions. As for the mass distributions, the $z_g$ spectra are different in the case of Pythia and Rope fragmentation, but the difference is not affected by the multiplicity. 

\begin{figure}[h!]
\centering
\includegraphics[width=0.4\textwidth]{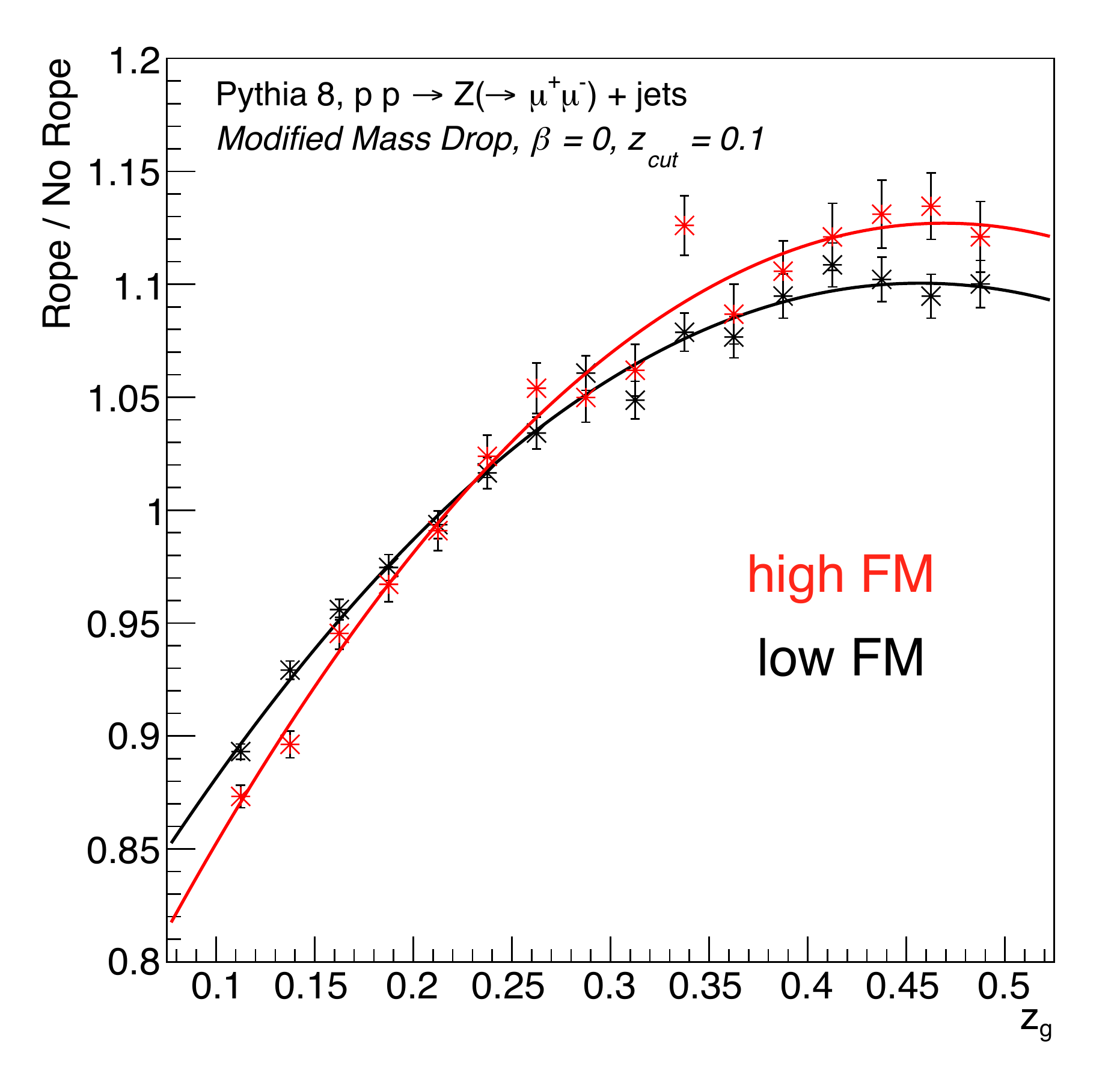}
\caption{The quadratic fits to the low- and high-multiplicity Rope/No Rope $z_g$ spectra shown for FM in Fig.~\ref{fig:softdrop}.}
\label{fig:fit}
\end{figure}

However, there is a small difference in the impact of the Rope hadronization on the effect of high multiplicity for the $z_g$ distribution.  The Rope hadronization predicts a lower multiplicity (Fig.~\ref{fig:multiplicity}) and thus more energy per particle.  This results in more jets that are completely removed from grooming (less pronounced second subjet).  However, given that there is a second subjet, the energy sharing is more equal as the energy is confined to fewer hadrons (thus they have higher energy per hadron).  Fitting the lower ratio panel of the FM $z_g$ distribution to a polynomial\footnote{We use a quadratic polynomial for $z_g > 0.1$; the $\chi^2/\text{NDF}\sim 1$ and does not significantly improve for higher order polynomials.}, as shown in Fig.~\ref{fig:fit}, results in a constant term that significantly differs within the MC statistics, which are about 30\% higher than the 7 TeV data statistics (the linear and quadratic terms also differ, but not as significantly).  The size of the effect is about 10\%.

\begin{figure}[h!]
\centering
\includegraphics[width=0.49\textwidth]{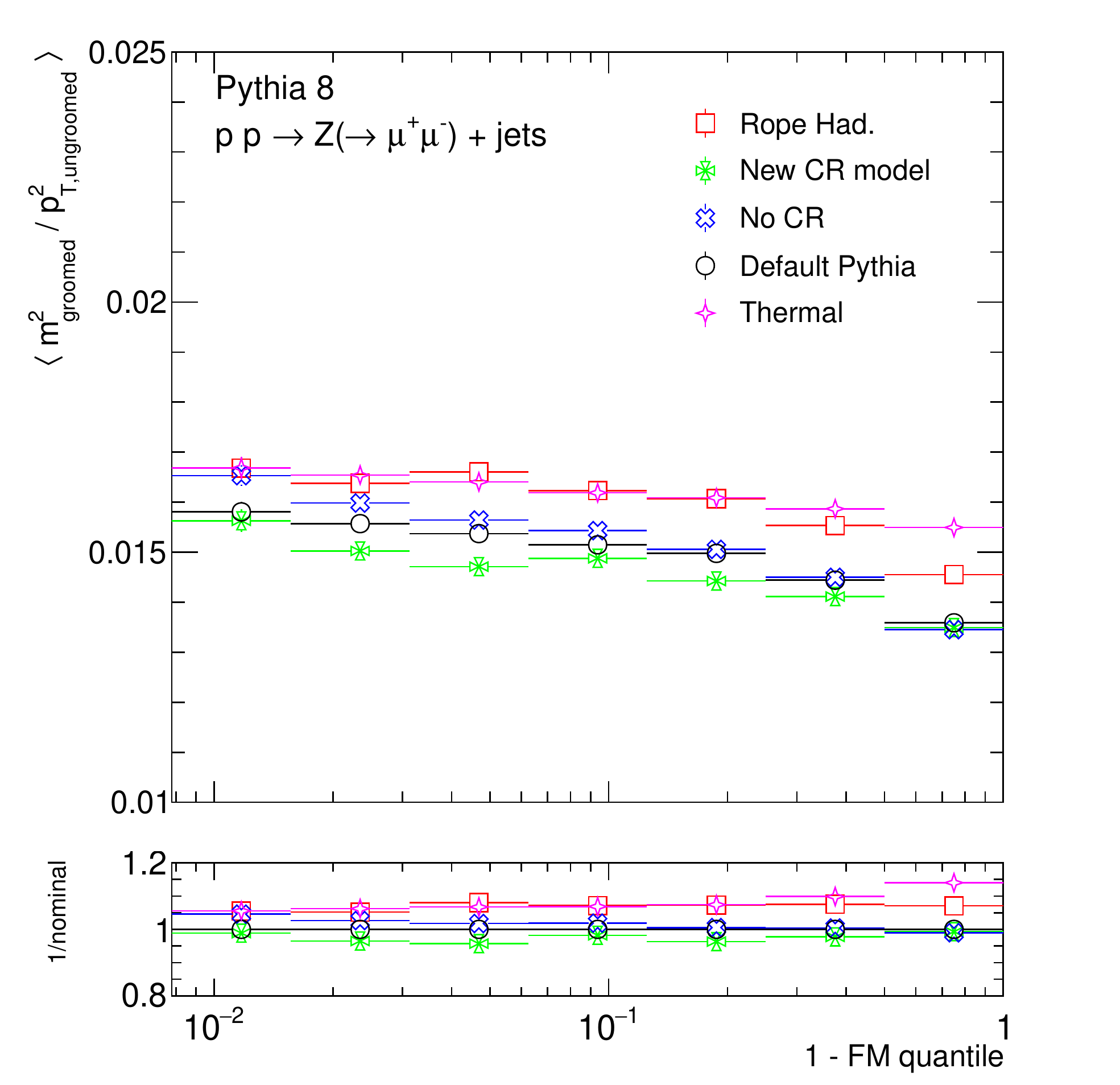}\includegraphics[width=0.49\textwidth]{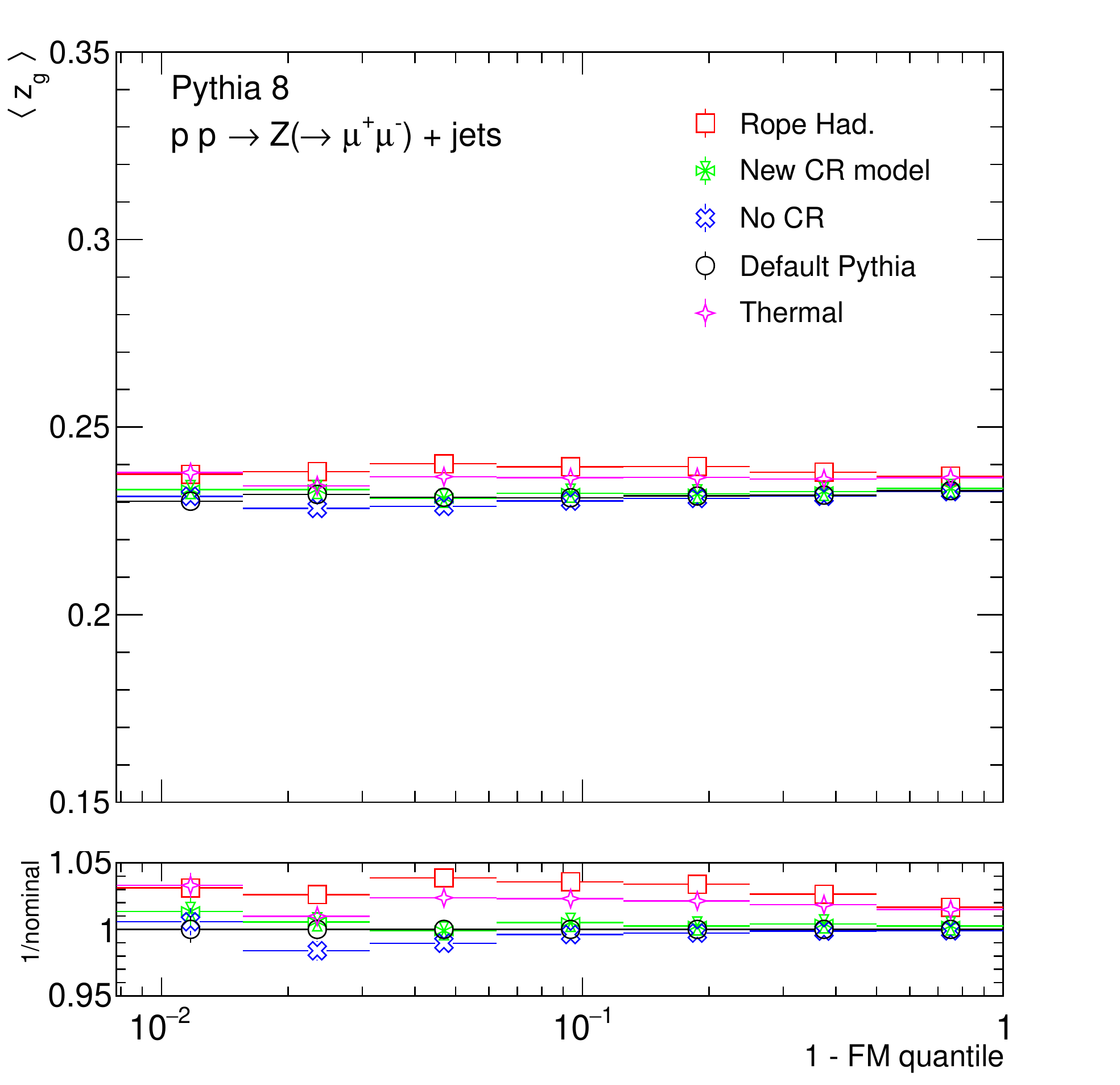}
\caption{The average mass (left) and $z_g$ (right) as a function of the FM multiplicity and various models of collective effects.  Higher multiplicities are to the left.   }
\label{fig:softdropmany}
\end{figure}

The additional models introduced in Sec.~\ref{sec:coleffects} are shown in Fig.~\ref{fig:softdropmany}.  The thermal model shows similar trends to the Rope hadronization for the mass, but the shift in the $z_g$ distribution is less pronounced than the Rope model.  The new CR model predicts little impact on the mass and $z_g$.

\begin{figure}[h!]
\centering
\includegraphics[width=0.495\textwidth]{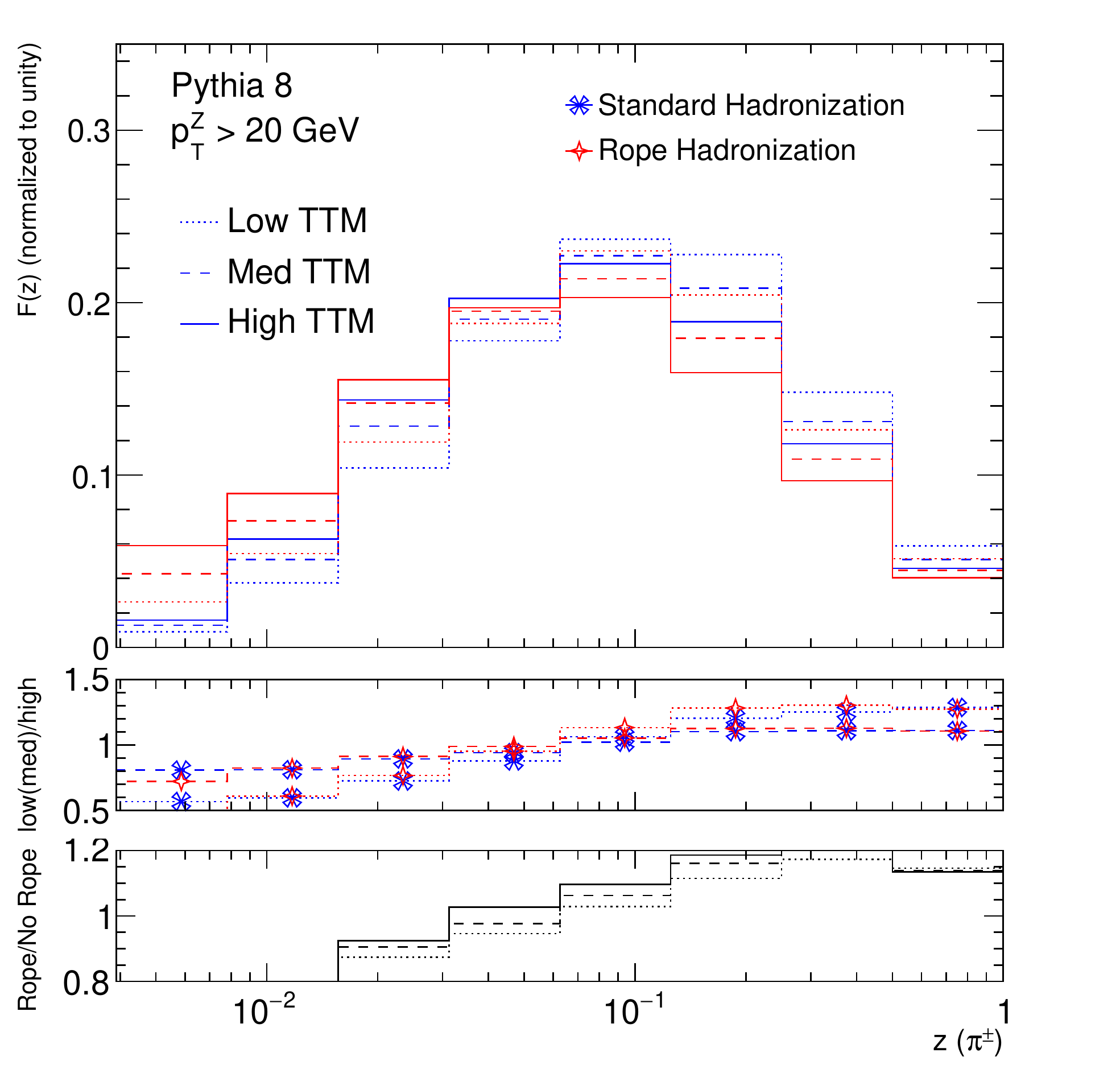}
\includegraphics[width=0.495\textwidth]{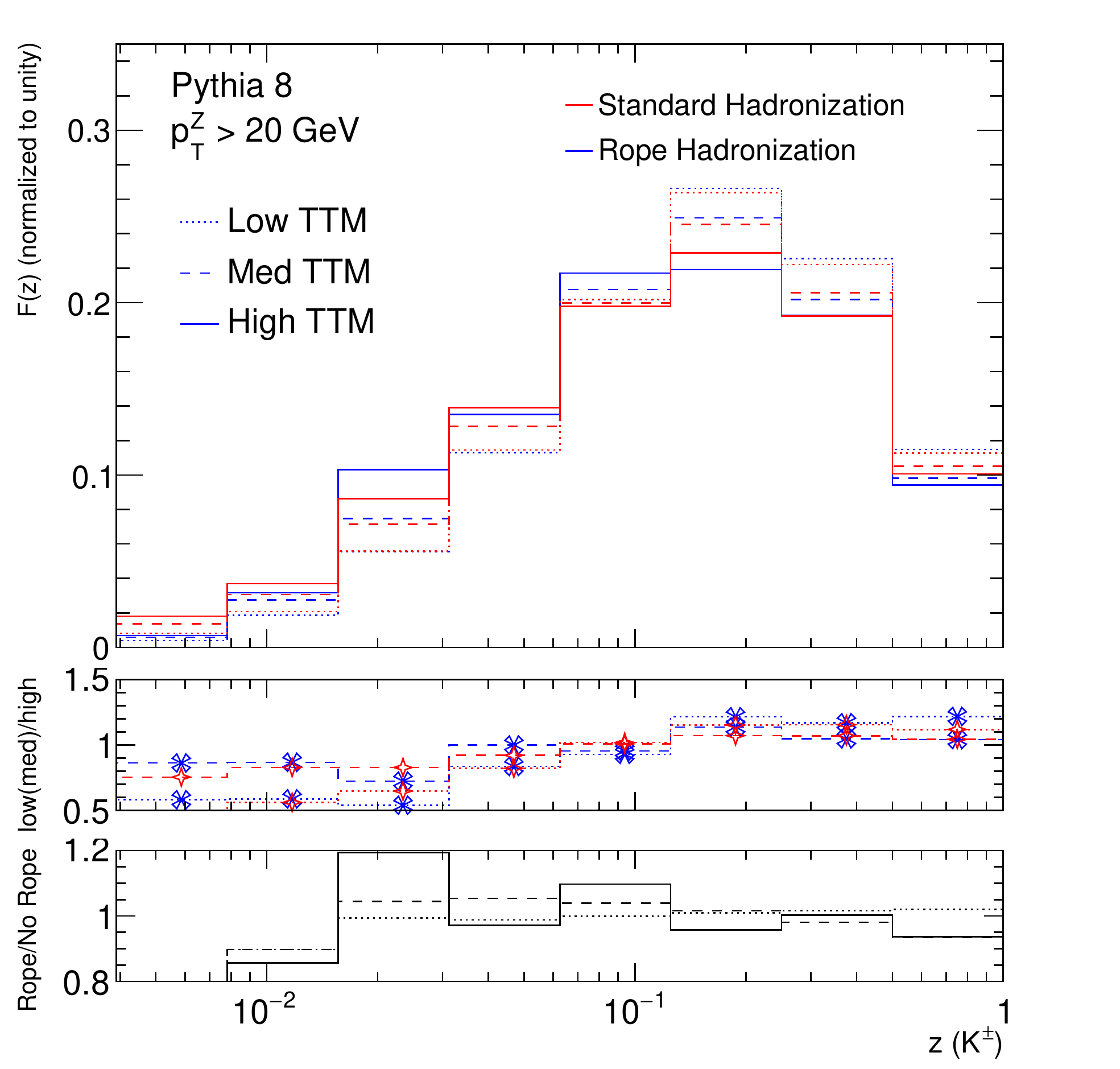}
\caption{The distribution of the momentum fraction ($z$) carried by pions (left) and kaons (right) for low ($<50$), medium ($50<\text{TTM}<100$), and high ($>100$) TTM for both the default and Rope hadronization models.  The middle panel shows the ratio of the distributions for high to low multiplicity for both hadronization models.  In the lower panel, the ratio between the Rope and standard hadronization models is displayed for all three multiplicity regions. }
\label{fig:fragfunc}
\end{figure}

Another well-studied jet substructure observable is the fraction of a jet's momentum carried by identified particles.  Figure~\ref{fig:fragfunc} shows this variant of the fragmentation function in various regions of TTM with and without the Rope hadronization model.  For both models, a higher multiplicity corresponds to a softer spectrum.  This is due in part to the increased multiplicity of UE that happens to fall in the jet catchment area.  Interestingly, there is a multiplicity-dependent difference in this effect between the Rope and standard hadronization models for pions, though it is less clear for kaons (due in part to limited MC statistics).

\clearpage

\section{Conclusions and Future Outlook}
\label{sec:concl}

Proton-proton interactions can be much more complex than vacuum parton-parton interactions.  When collided with significant overlap, collective effects observed also in more extended systems are suggestive of a common origin.  Using MC simulations of models with collective effects, we have studied observables related to jets that may be sensitive to the source of the observed phenomena.   No one model can explain all of the observed collective effects, but they provide useful benchmarks for probing quenching-like behavior.  Interpreting the strangeness enhancement in high multiplicity $pp$ as a sign of a QGP, it will be interesting to next try to quantify the expected size of such a QGP by scaling up to HI and then predicting the magnitude of potential jet quenching.  We note that several studies have appeared in the literature, discussing the possibility of a QGP formation in $pp$ and pA collisions~\cite{Tywoniuk:2014hta,Zakharov:2013gya,Chen:2015qmd,Kang:2015mta}. Their conclusions and estimates of the properties of this QGP state vary significantly, and we are not in the position to project in a robust way their impact on the jet quenching effects that could arise in the context of the $Z$+jet observables discussed here. This would require setting up an event simulation framework incorporating the possible development of a QGP in $pp$ collisions, a task that goes beyond the scope of our simple study, and which will hopefully be picked up by more expert colleagues. 

Independently of the possible quenching effects induced by a mini-QGP, our study shows some interesting features of the Rope fragmentation, and differences with respect to the standard Pythia fragmentation at the level of several percent, when considering jet-related quantities, such as the groomed mass and the $z_g$ spectra. The experimental study of these quantities can therefore provide additional handles in the tuning of these alternative fragmentation models, or in the development of new ones. 
Measurements with $Z/\gamma$+jets should be possible with high precision using ATLAS and CMS and the observables and trends presented here provide a baseline for a full experimental investigation.

\section{Acknowledgements}
\label{sec:thanks}

We would like to thank the organizers and participants of the ``Collective effects in small collisions systems'' workshop at CERN for stimulating discussions.  In addition, we are grateful to Jesse Thaler and Peter Jacobs for useful discussions and for recommending the study of jet substructure observables and forward multiplicity, respectively.  This work was supported in part by the Office of High Energy Physics of the U.S. Department of Energy under contract DE-AC02-05CH11231.

\clearpage

\appendix

\section{Parameter Settings}
\label{sec:params}

Table~\ref{tab:params} describes all of the simulation parameters used in this paper.

\begin{table}[h!]
\centering
\noindent\adjustbox{max width=\textwidth}{
\begin{tabular}{ |c|c|c|c|c|c |}
 \hline
Parameter & Default Pythia & Rope & New CR & No CR & Thermal Model\\
 \hline
 \texttt{StringPT:sigma}   & 0.335    & 0.31 & -- & -- & --  \\
\texttt{StringZ:aLund}   & 0.68    & 0.38 & 0.36 & -- & --  \\    
  \texttt{StringZ:bLund}   & 0.98    & 0.37 & 0.56 & -- & --  \\
  \texttt{StringFlav:probStoUD}   & 0.19    & 0.21 & 0.2 & -- & --  \\ 
  \texttt{StringFlav:probSQtoQQ}   & 1.0    & 0.915 & -- & -- & --  \\ 
  \texttt{StringFlav:probQQ1toQQ0}   & 0.027    & 0.0275 & -- & -- & --  \\ 
    \texttt{StringFlav:probQQ1toQQ0join}   & 0.5,0.7,0.9,1.0    & -- & 0.0275,0.0275,0.0275,0.0275 & -- & --  \\ 
\texttt{StringFlav:probQQtoQ}   & 0.09    & 0.073 & 0.078 & -- & --  \\ 
\texttt{StringZ:aExtraDiquark}   & 0.5    & 0.97 & -- & -- & --  \\ 
\texttt{RadiusRatio}   & N/A    & 0.2 & -- & -- & --  \\ 
\texttt{RapiditySpan}   & N/A    & 0.5 & -- & -- & --  \\ 
\texttt{MultiPartonInteractions:pT0Ref}   & 2.15    & -- & -- & -- & 2.5  \\ 
\texttt{BeamRemnants:remnantMode}   & 0    & -- & 1 & -- & --  \\ 
\texttt{ColourReconnection:reconnect}   & on    & -- & -- & off & --  \\ 
\texttt{ColourReconnection:mode}   & 0    & -- & 1 & -- & --  \\ 
\texttt{ColourReconnection:range}   & 1.8    & -- & -- & -- & 1.1  \\ 
\texttt{ColourReconnection:allowDoubleJunRem}   & on    & -- & off & -- & --  \\ 
\texttt{ColourReconnection:m0}   & N/A    & -- & 0.3 & -- & --  \\ 
\texttt{ColourReconnection:allowJunctions}   & N/A    & -- & on & -- & --  \\ 
\texttt{ColourReconnection:junctionCorrection}   & N/A    & -- & 1.2 & -- & --  \\ 
\texttt{ColourReconnection:timeDilationMode}   & N/A    & -- & 2 & -- & --  \\ 
\texttt{ColourReconnection:timeDilationPar}   & N/A    & -- & 0.18 & -- & --  \\ 
\texttt{StringPT:thermalModel}   & off    & -- & -- & -- & on  \\ 
\texttt{StringPT:temperature}   & N/A    & -- & -- & -- & 0.21  \\ 
\texttt{StringFlav:BtoMratio}   & N/A    & -- & -- & -- & 0.357  \\ 
\texttt{StringFlav:StrangeSuppression}   & N/A    & -- & -- & -- & 0.5  \\ 
\texttt{StringPT:expNSP}   & N/A    & -- & -- & -- & 0.13  \\ 
\texttt{HadronLevel:HadronScatter}   & off    & -- & -- & -- & on  \\ 
\texttt{HadronScatter:mode}   & N/A    & -- & -- & -- & 0  \\ 
\texttt{HadronScatter:maxProbDS}   & N/A    & -- & -- & -- & 0.5  \\ 
 \hline
 \hline
 Reference: & ~\cite{Skands:2014pea} & ~\cite{Bierlich:2016faw} & ~\cite{Christiansen:2015yqa} & & ~\cite{Fischer:2016zzs}\\
 \hline
\end{tabular}}
\label{tab:params}
\caption{The parameters used for the various models described in this paper.  A `--' indicates that the same value as the default model is used.}
\end{table}

\clearpage

\section{Plots with Alternative Multiplicity Definitions}
\label{sec:altmult}

\begin{figure}[h!]
\centering
\includegraphics[width=0.48\textwidth]{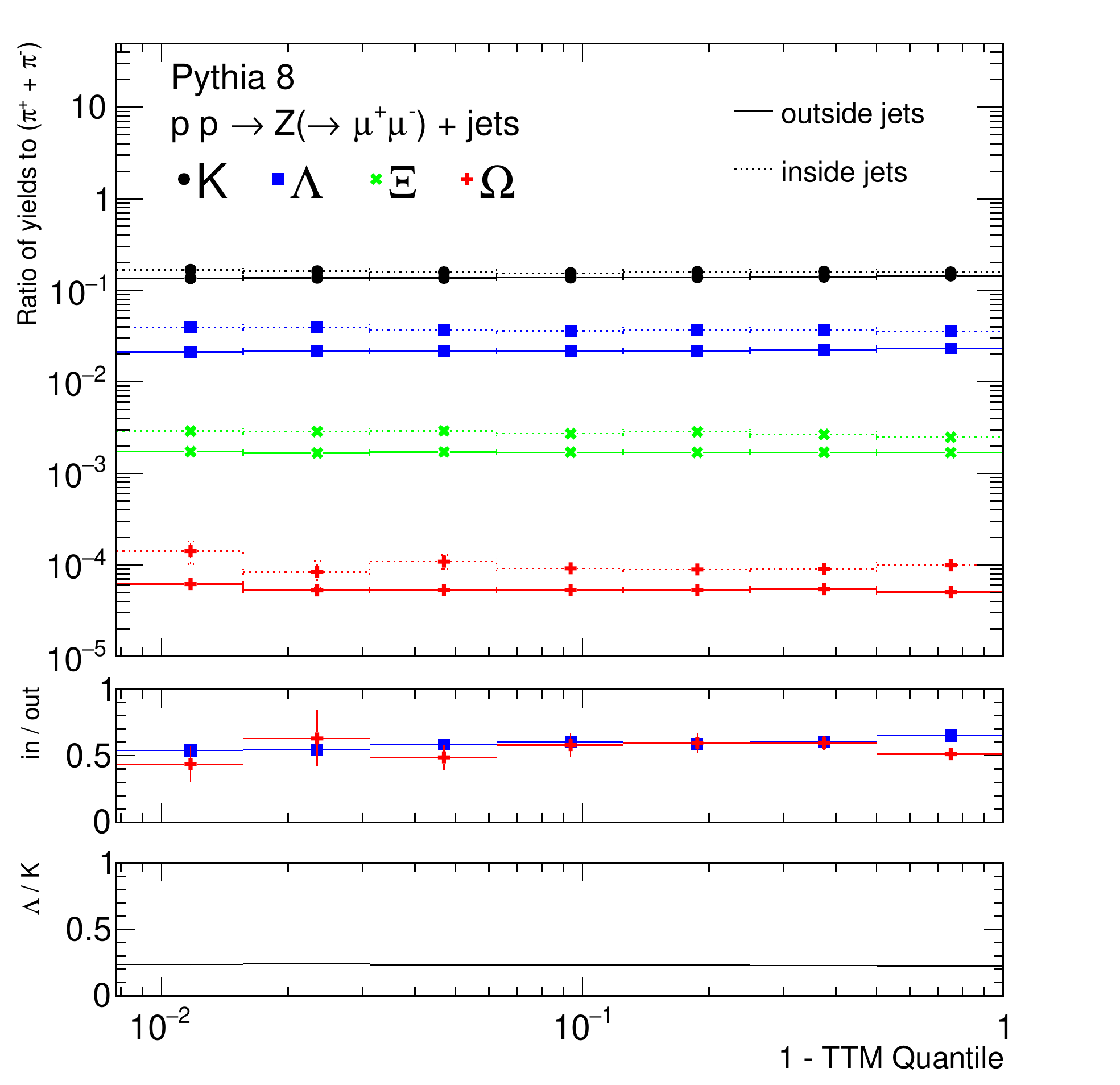}\includegraphics[width=0.48\textwidth]{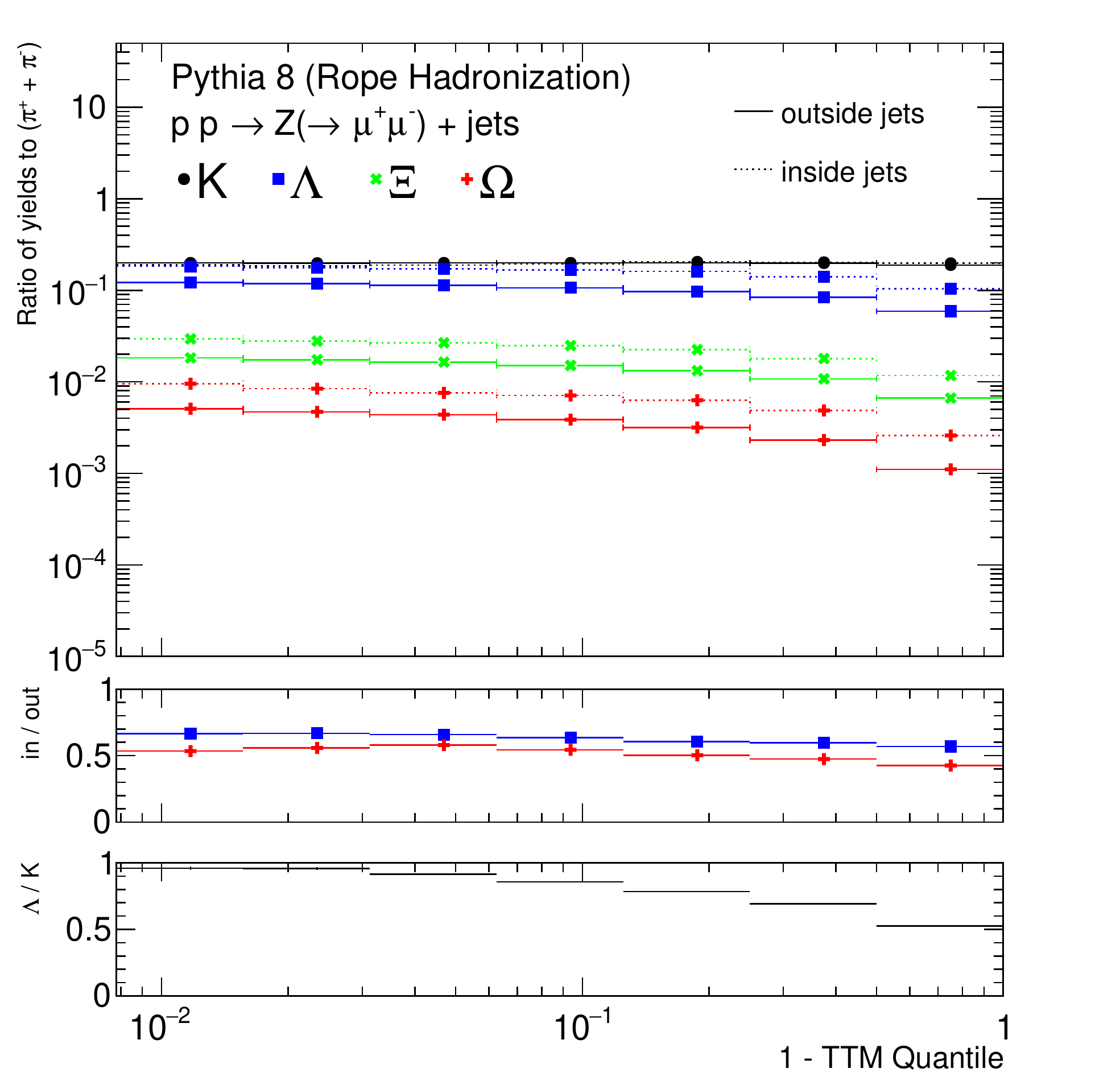}
\caption{Same as Fig.~\ref{fig:strangeTTM}, but as a function of the TTM quantile.}
\label{fig:strangeFM}
\end{figure}

\begin{figure}[h!]
\centering
\includegraphics[width=0.48\textwidth]{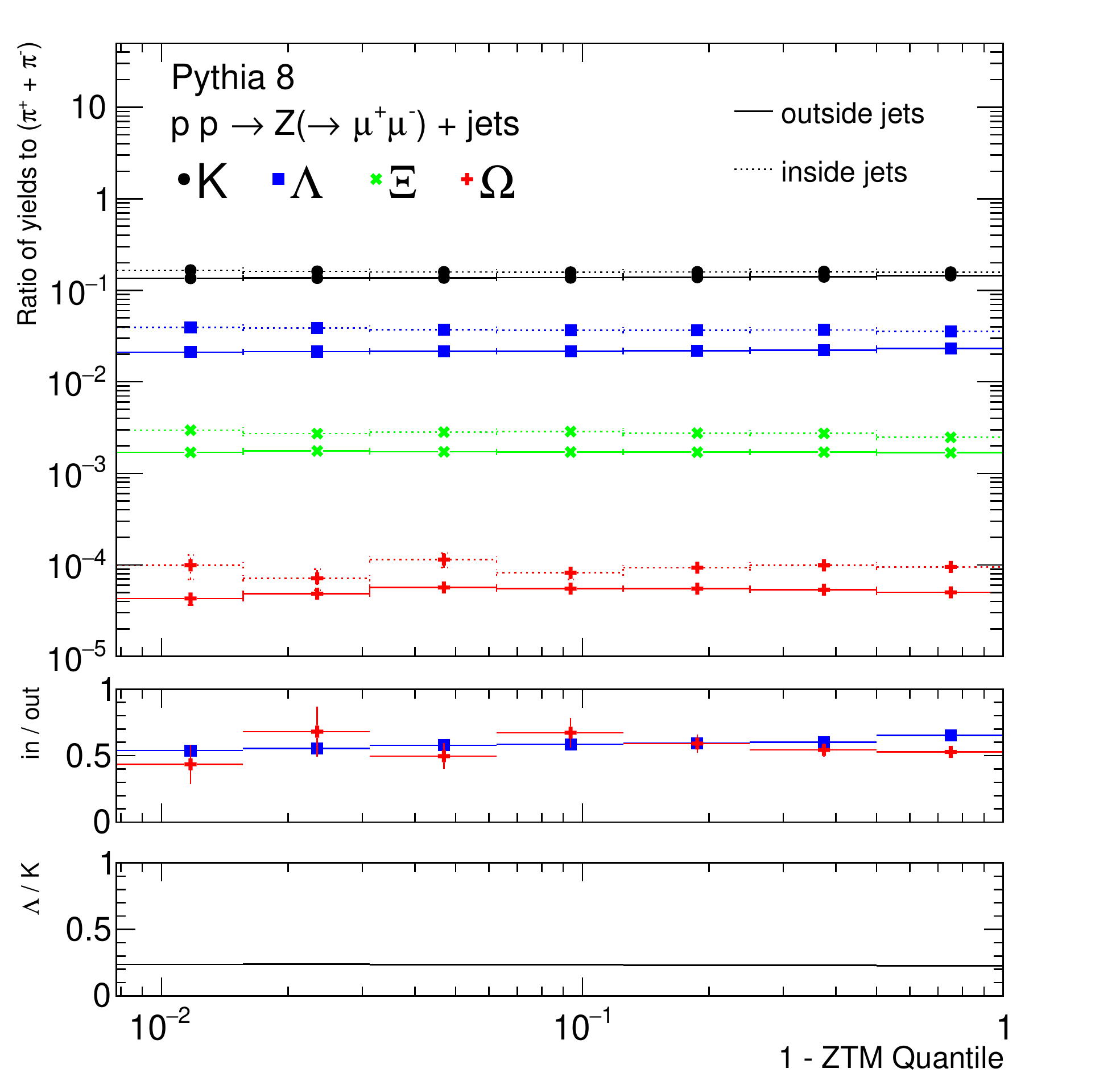}\includegraphics[width=0.48\textwidth]{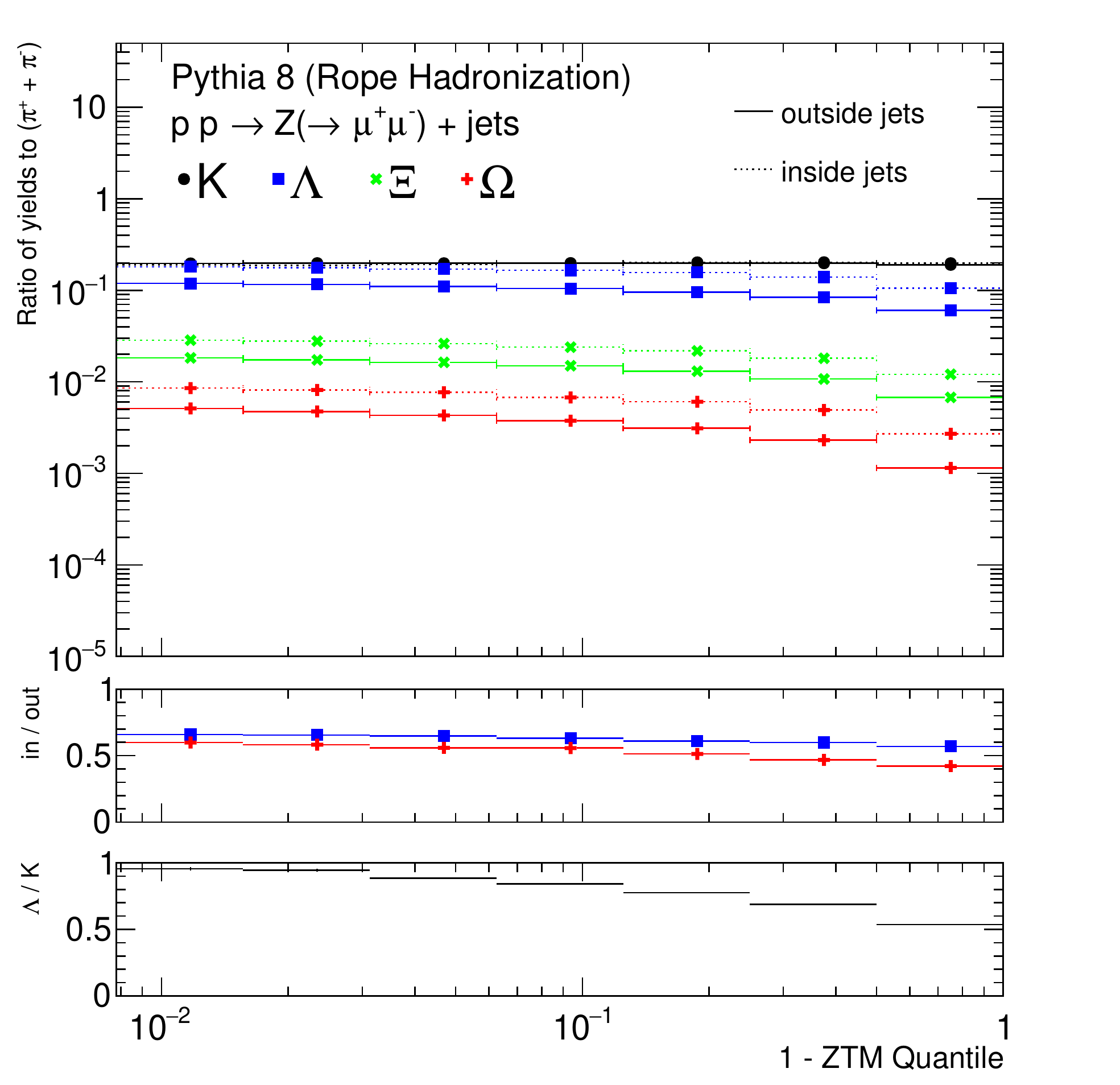}
\caption{Same as Fig.~\ref{fig:strangeTTM}, but as a function of the ZTM quantile.}
\label{fig:strangeZTM}
\end{figure}

Figures~\ref{fig:strangeZTM} and~\ref{fig:strangeFM} show plots of strangeness enhancement for the TTM and ZTM multiplicity definitions. 

\begin{figure}[h!]
\centering
\includegraphics[width=0.4\textwidth]{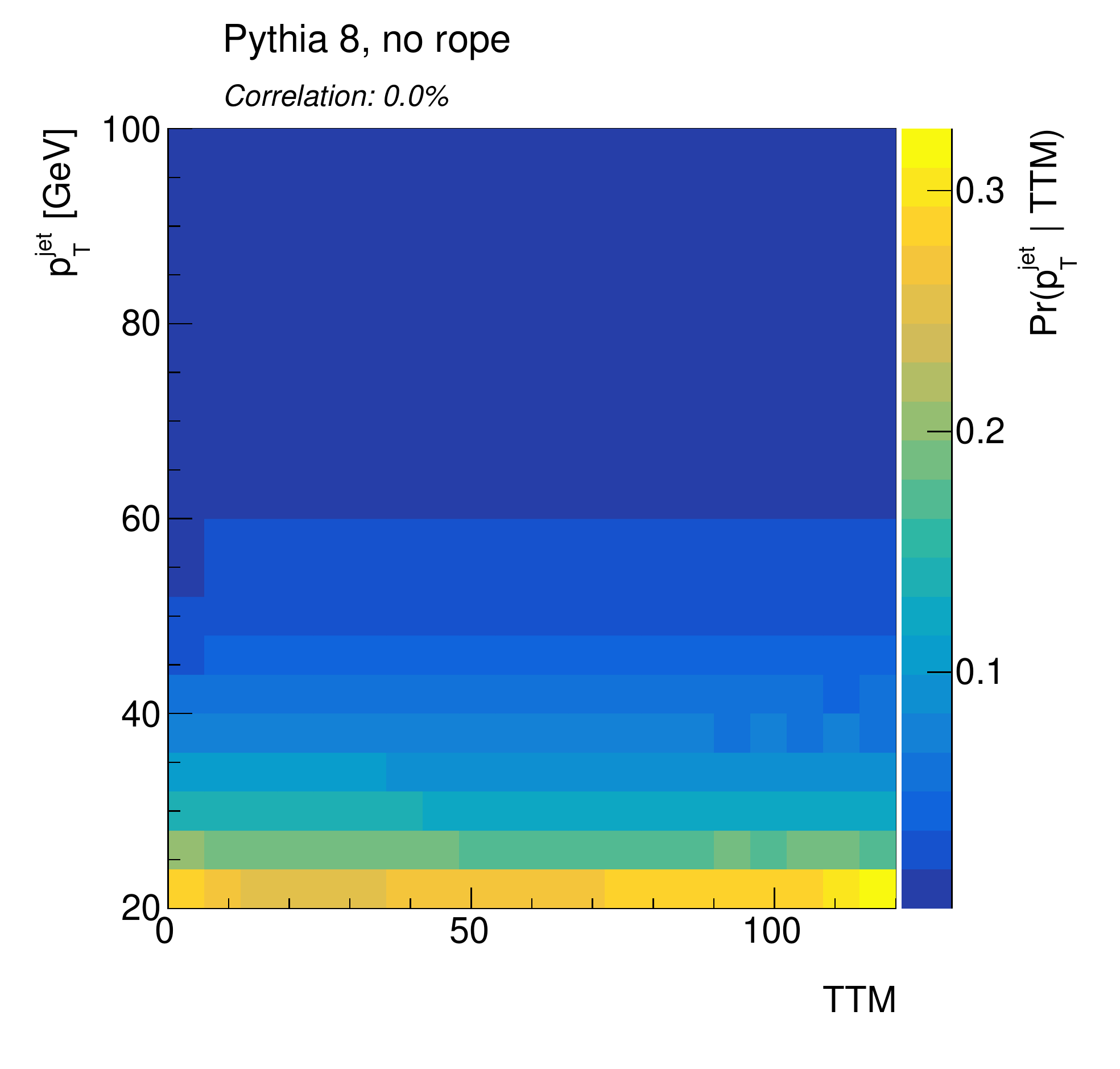}\includegraphics[width=0.4\textwidth]{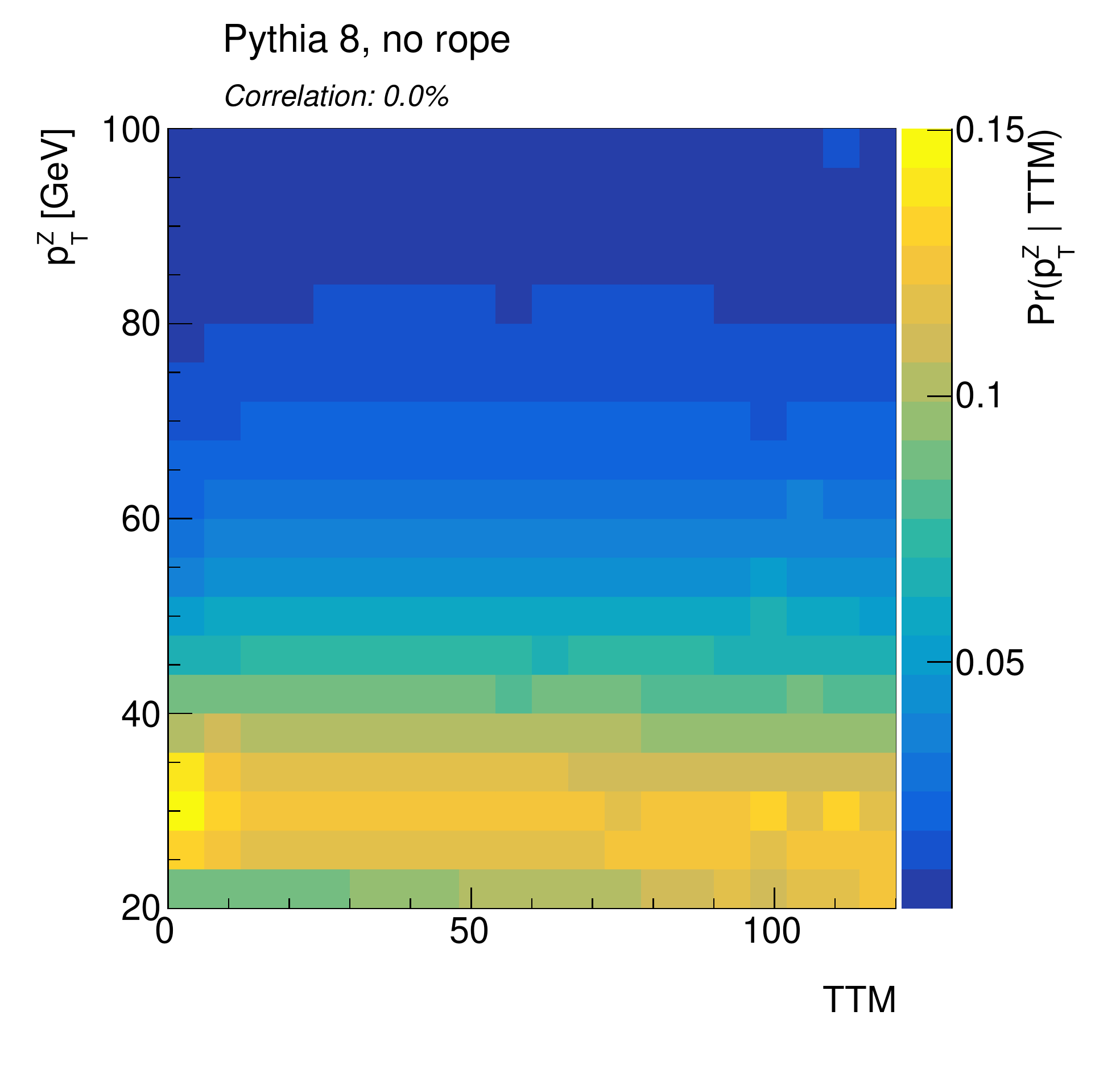}
\includegraphics[width=0.4\textwidth]{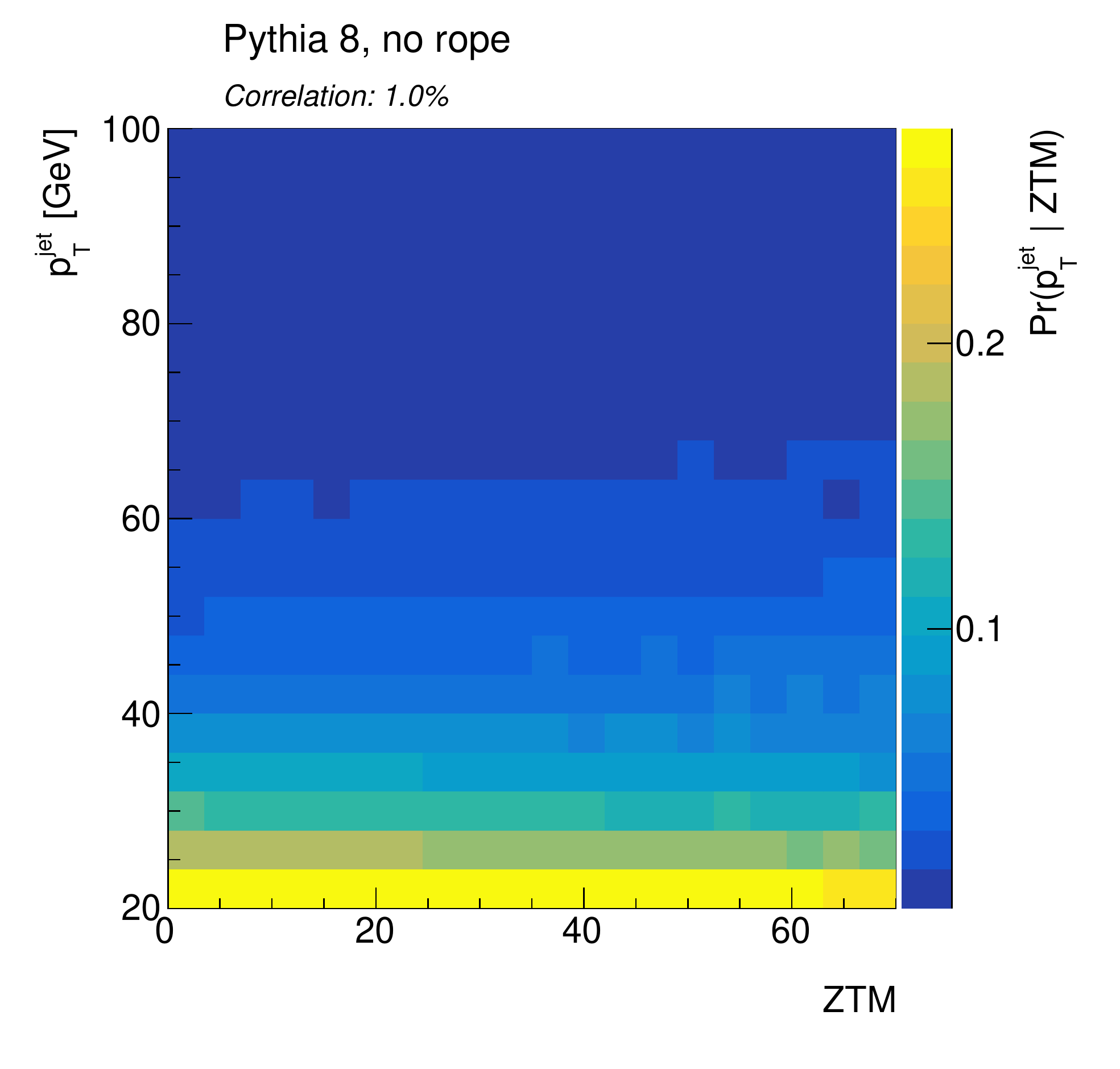}\includegraphics[width=0.4\textwidth]{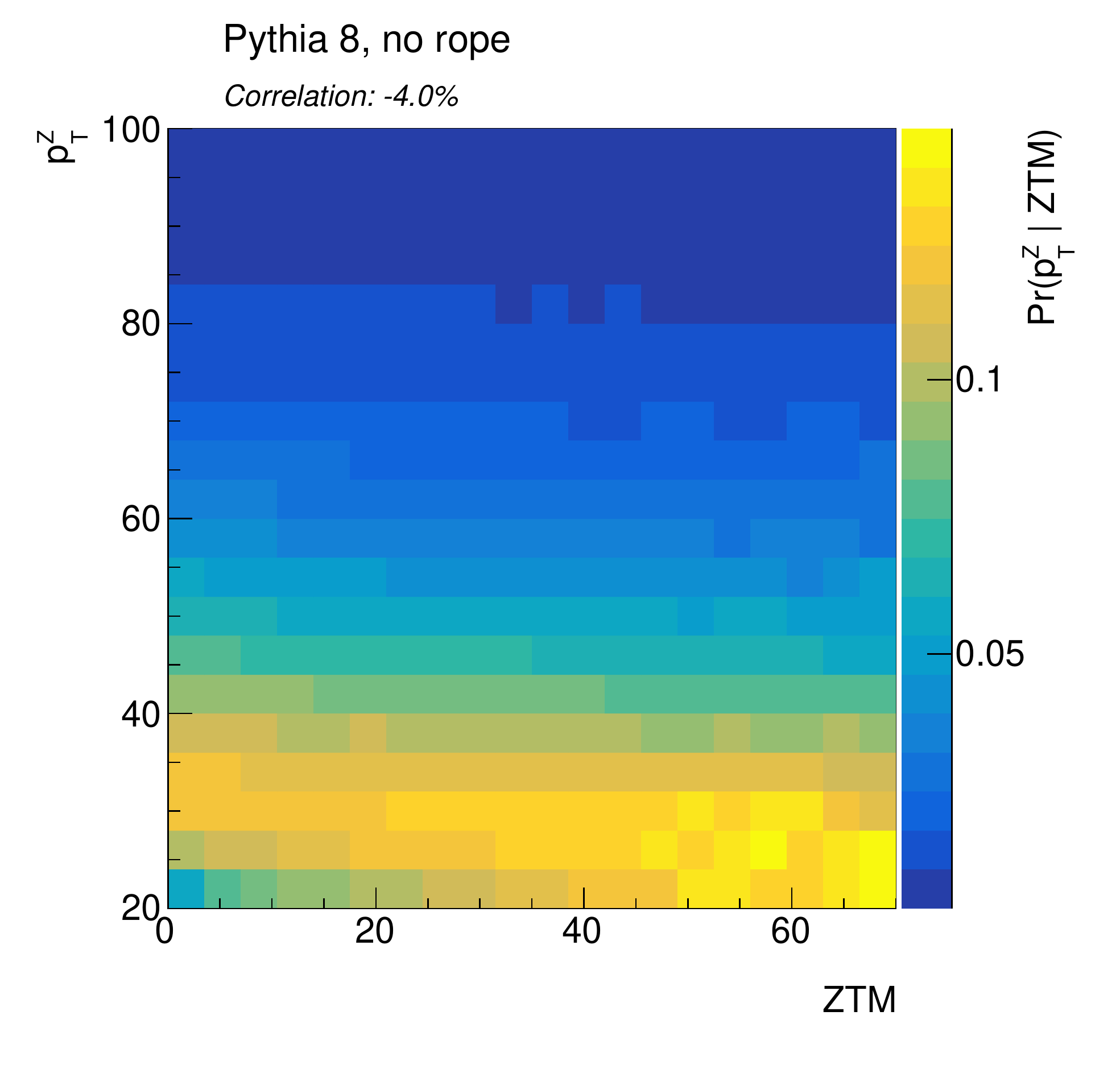}
\caption{The distribution of the jet (left) or $Z$ boson (right) $p_\text{T}$ given the event multiplicity defined by TTM (top) and ZTM (bottom).  The linear correlation coefficient is presented at the top of each plot.}
\label{fig:correlationptmultapp}
\end{figure}

\begin{figure}[h!]
\centering
\includegraphics[width=0.48\textwidth]{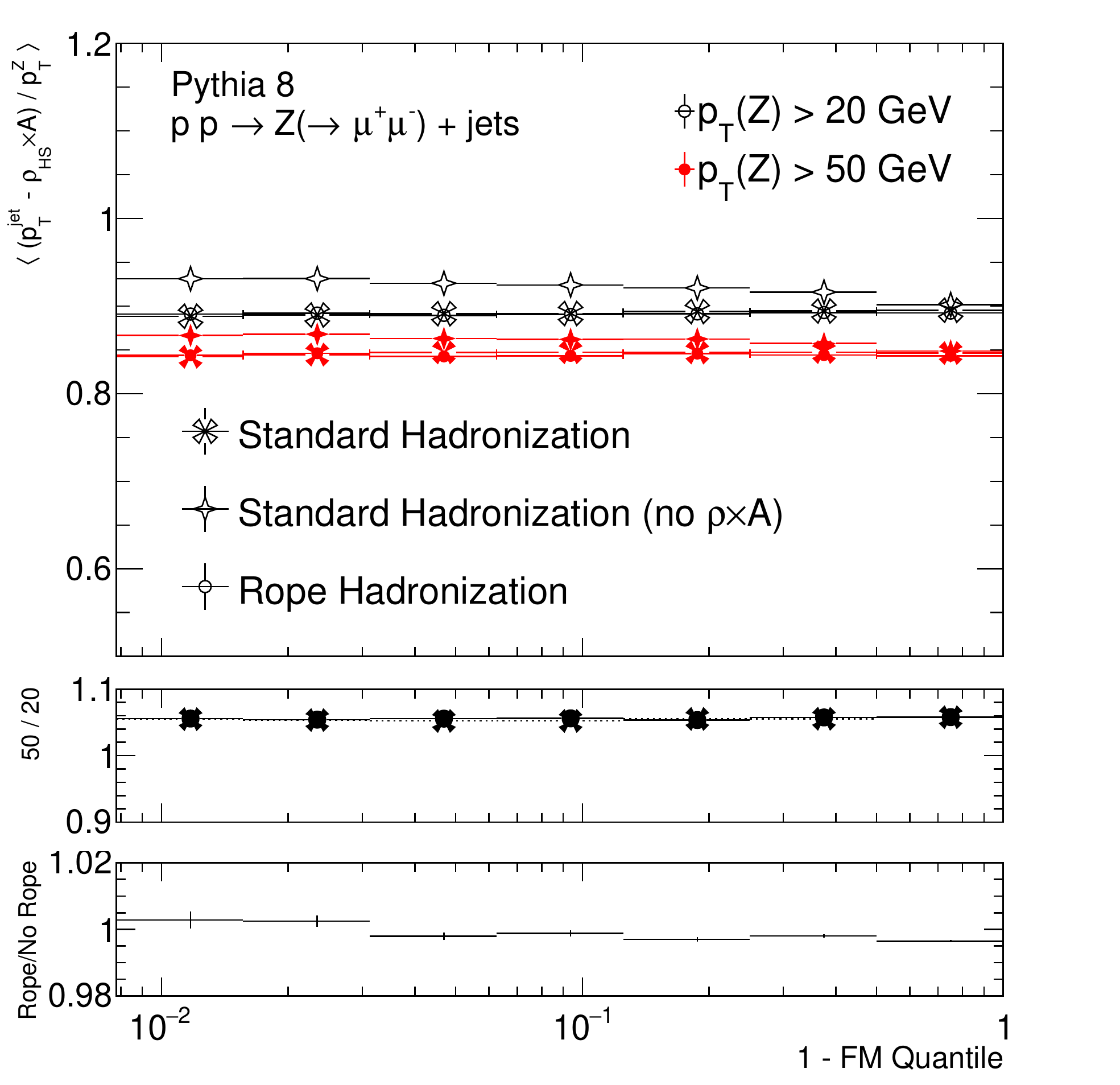}
\caption{The average fractional transverse momentum imbalance between leading jet and $Z$ boson, $x_{ZJ}=p_\text{T,$J$} / p_\text{T,$Z$}$,  as a function of track multiplicity quantile, for TTM (left) and ZTM (right).  Three curves in the upper panels indicate the values for the standard hadronization with and without an area correction as well as the rope hadronization with the correction.  Higher multiplicities are to the left.  The middle panel shows the ratio of the higher to lower $Z$ boson $p_\text{T}$ cut (standard hadronization is shown with a dashed line, rope with solid), both including the areas correction.  Finally, the lower panel shows the ratio between the rope and standard hadronization models with the $p_\text{T,$Z$}>20$ GeV requirement and both with the areas correction.}
\label{fig:jetsapp}
\end{figure}

\begin{figure}[h!]
\centering
\includegraphics[width=0.49\textwidth]{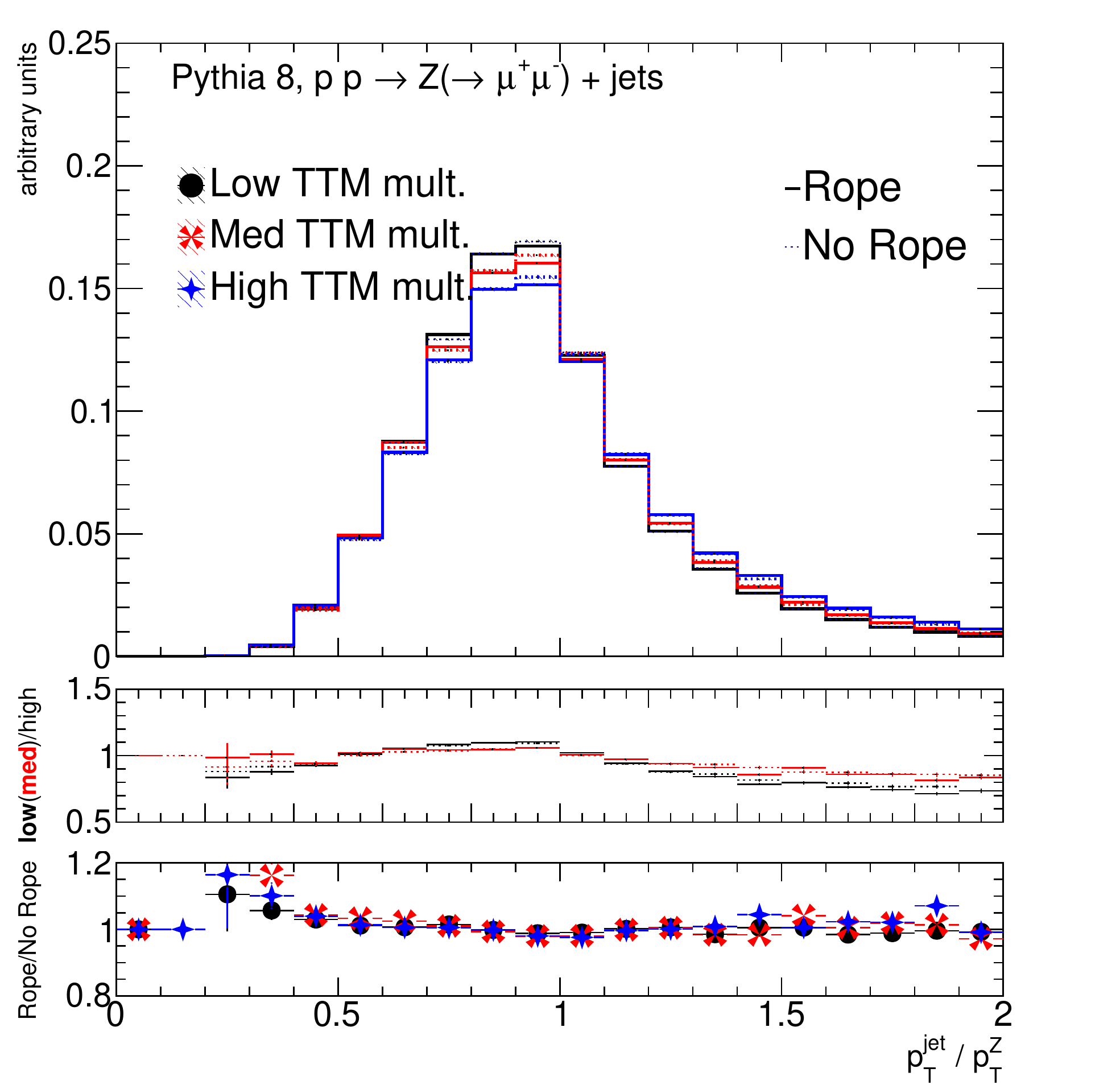}\includegraphics[width=0.49\textwidth]{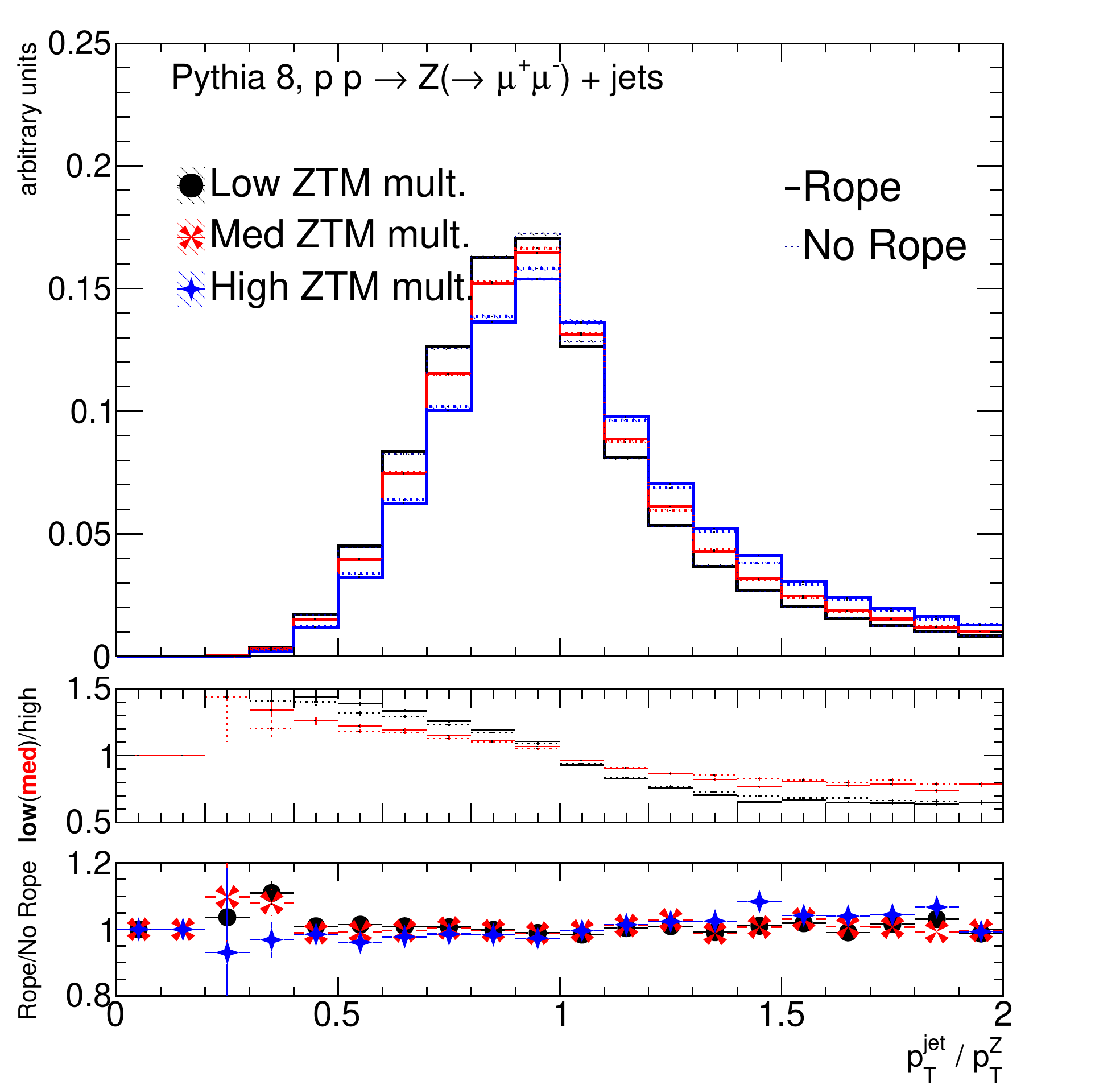}
\caption{The distribution of the ratio $x_{ZJ}=p_\text{T,$J$} / p_\text{T,$Z$}$ for three bins of the TTM (left) and ZTM (right) multiplicity: the 50th, 75th, and 90th percentile.  The middle panel shows the ratio of the distribution for high to low multiplicity for both hadronization models (standard hadronization with a dotted line).  In the lower panel, the ratio between the Rope and standard hadronization models is displayed for all three multiplicity regions.}
\label{fig:ptratioapp}
\end{figure}

\begin{figure}[h!]
\centering
\includegraphics[width=0.4\textwidth]{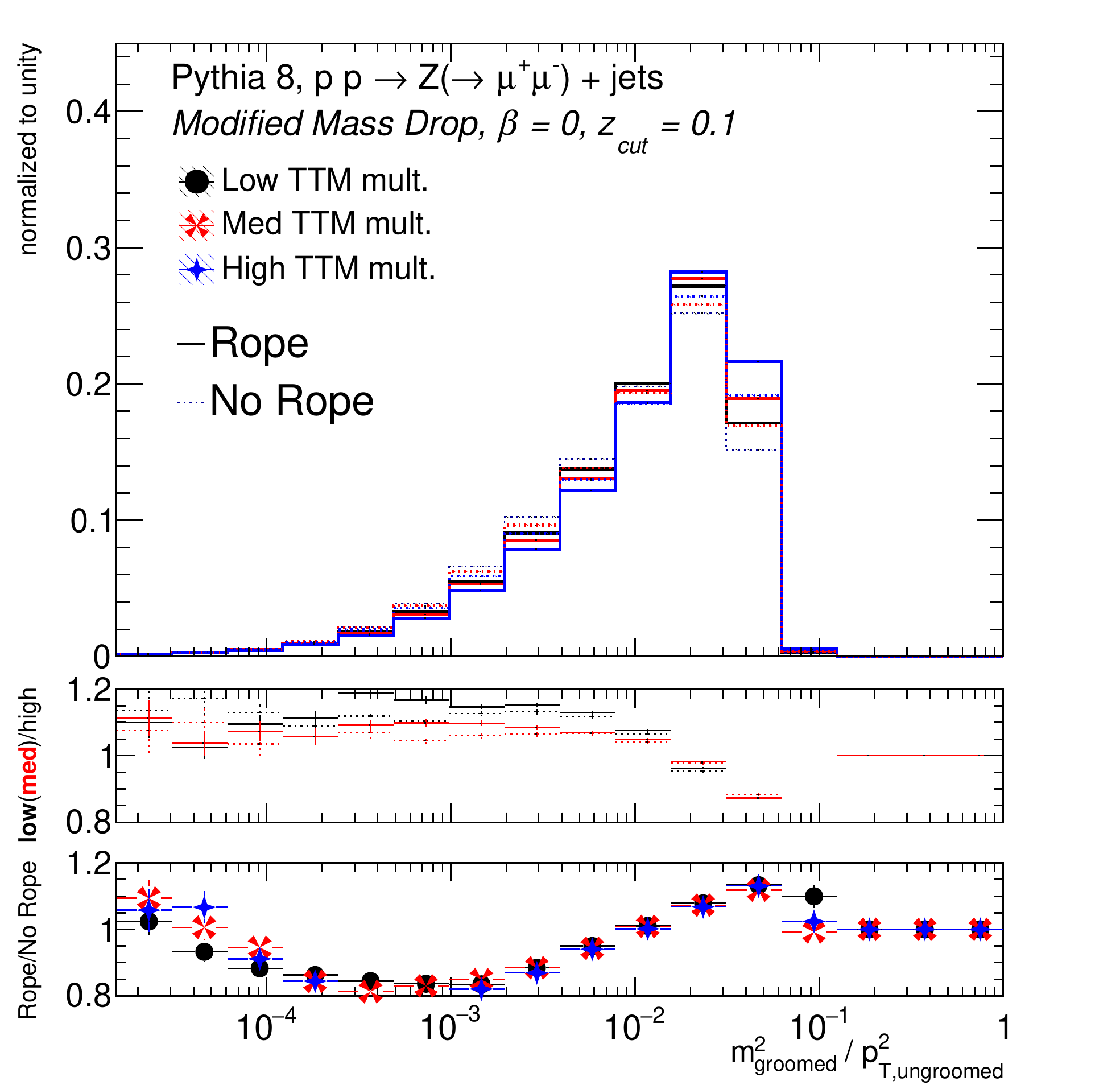}
\includegraphics[width=0.4\textwidth]{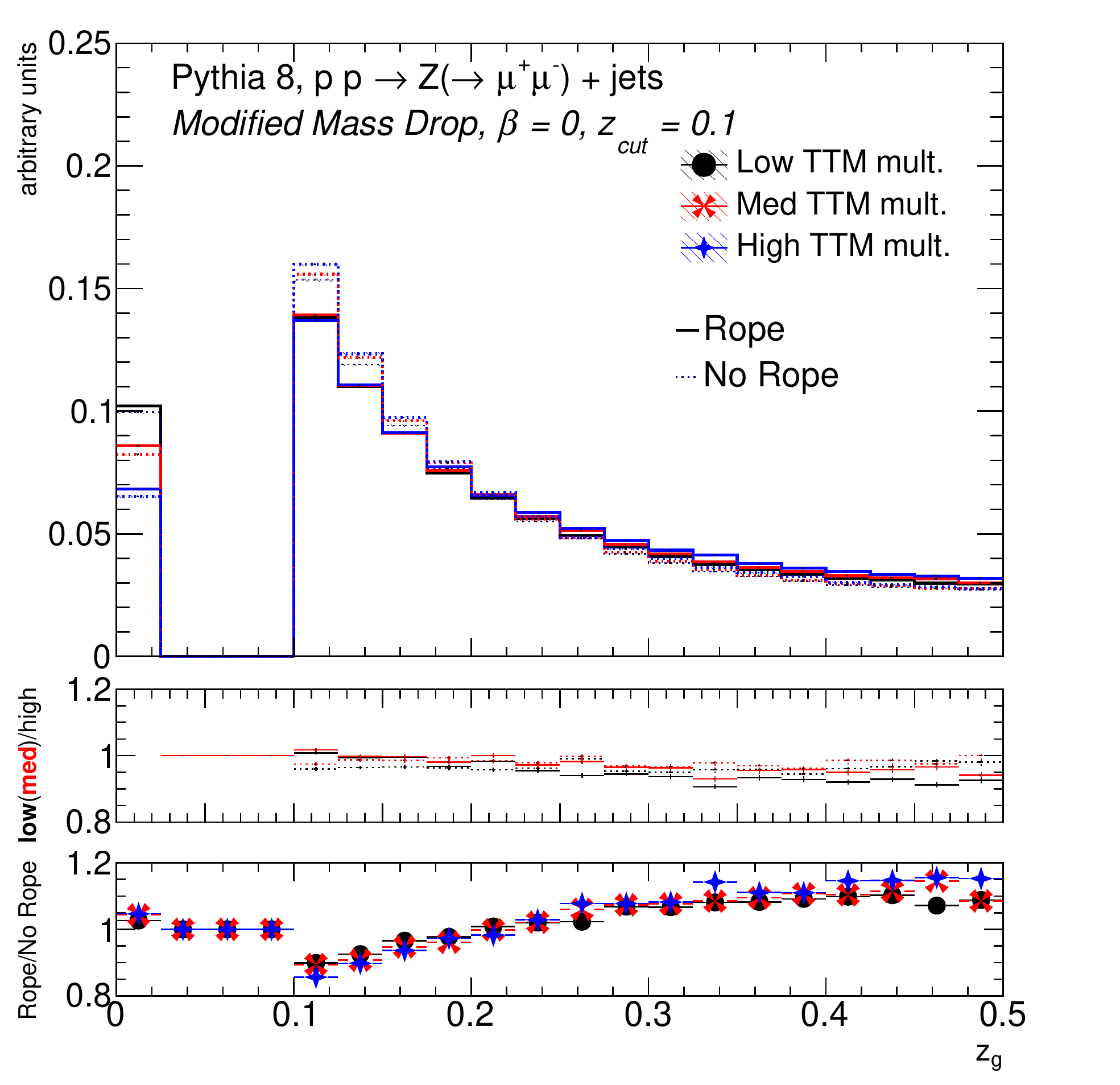}
\includegraphics[width=0.4\textwidth]{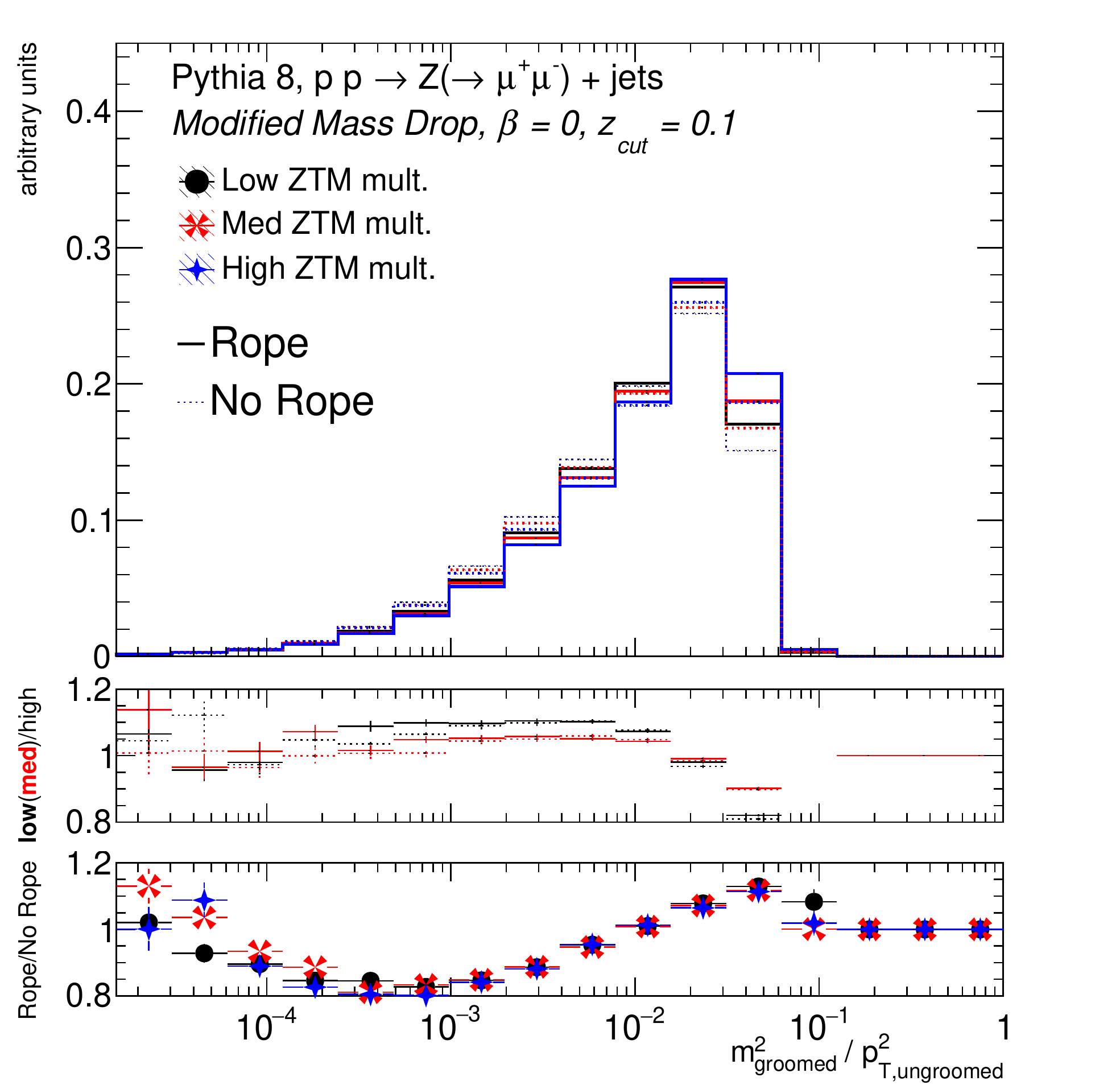}
\includegraphics[width=0.4\textwidth]{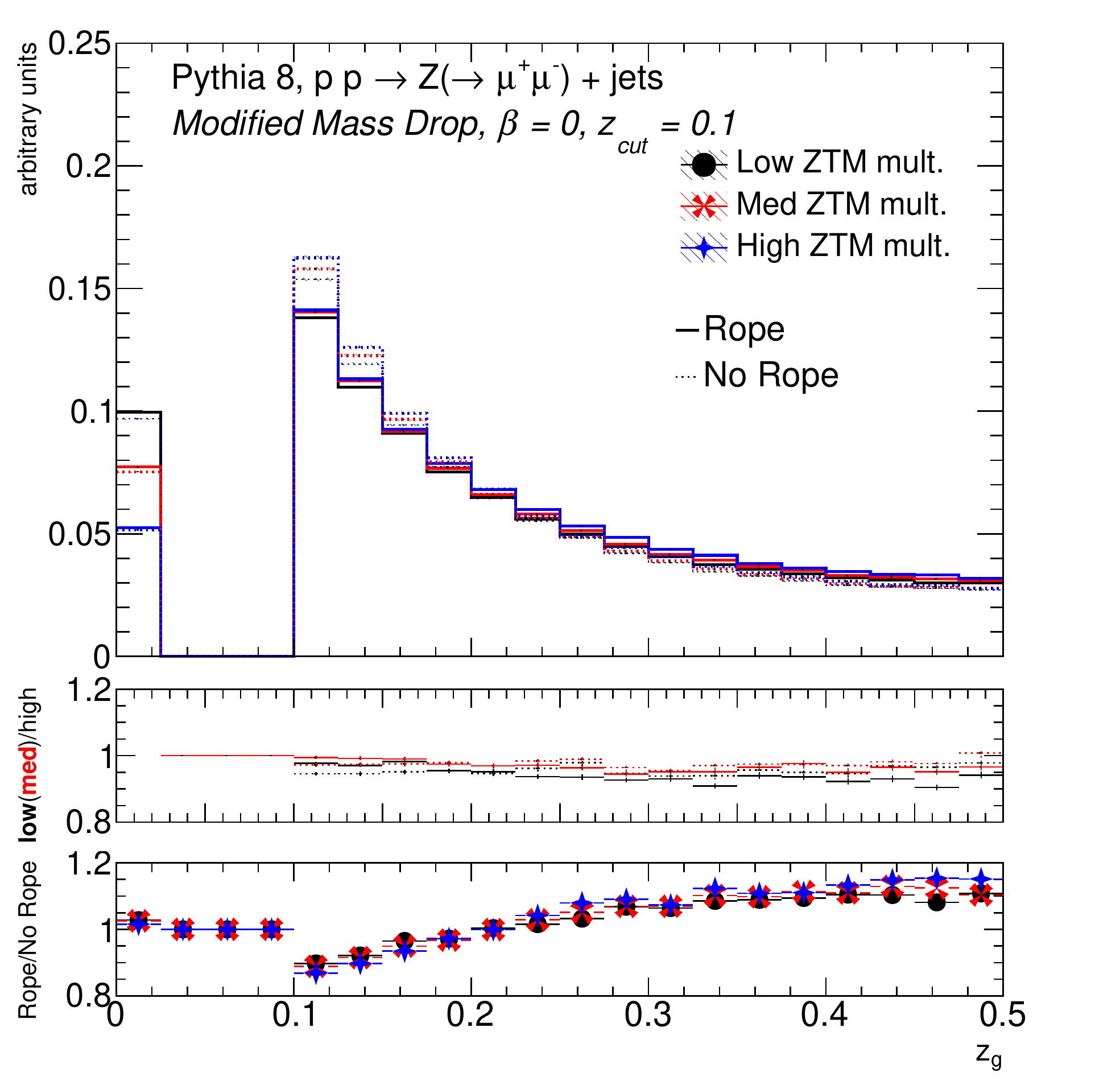}
\caption{Modified Mass Drop (also Soft Drop with $\beta=0$) jet mass (left) and momentum sharing $z_g$ (right) for three bins in FM that correspond to the 50th, 75, and 90th percentiles.  The middle panel shows the ratio of the distributions for high to low multiplicity for both hadronization models (standard hadronization with a dotted line).  In the lower panel, the ratio between the Rope and standard hadronization models is displayed for all three multiplicity regions.  Due to the algorithm value $z_\text{cut}=0.1$, $z_g \geq 0.1$; when the entire jet is groomed away, $z_g=0$.  The top plots use TTM for the multiplicity while the bottom plots use ZTM.}
\label{fig:softdropapp}
\end{figure}

\clearpage

\bibliographystyle{report}
\bibliography{myrefs-mlm}{}

\end{document}